\documentclass{aa}
\usepackage{rotating}
\usepackage{graphics}
\usepackage{times}
\def\kms{km\,s$^{-1}$} 
\def\Am{\AA\,mm$^{-1}$} 
\def\hb{$\rm H\beta$}
\def\ha{$\rm H\alpha$}
\def\nii{\lbrack\ion{N}{ii}\rbrack}
\def\oiii{\lbrack\ion{O}{iii}\rbrack}
\def\feii{\ion{Fe}{ii}}
\def\feiii{\lbrack\ion{Fe}{iii}\rbrack}
\def\fevii{\lbrack\ion{Fe}{vii}\rbrack}
\def\fex{\lbrack\ion{Fe}{x}\rbrack}
\def\fexiv{\lbrack\ion{Fe}{xiv}\rbrack}
\def\neiii{\lbrack\ion{Ne}{iii}\rbrack}

\def\cai{\ion{Ca}{i}}
\def\caii{\ion{Ca}{ii}}
\def\mgi{\ion{Mg}{i}}
\def\oi{\lbrack\ion{O}{i}\rbrack}
\def\oii{\lbrack\ion{O}{ii}\rbrack}
\def\roiii{$\lambda$5007$/$H$\beta$}
\def\rheii{$\lambda$4686$/$H$\beta$}
\def\rnii{$\lambda$6583$/$H$\alpha$}
\def\roi{$\lambda$6300$/$H$\alpha$}
\def\rliner{$\lambda$6300$/\lambda$5007}
\def\hii{\ion{H}{ii}}
\def\heii{\ion{He}{ii}}
\def\hi{\ion{H}{i}}
\begin{document}
\thesaurus{03(11.01.2;   
              11.14.1)}  
\title{AGNs with composite spectra
\thanks{{\scriptsize Based on observations collected at the 
Observatoire de Haute-Provence 
(CNRS), France, and {\it Hubble Space Telescope (HST)} data obtained from the 
Space Telescope European Coordinating Facility (ST-ECF) archive. Tables 5 
and 6 are also available in electronic form at the CDS via anonymous ftp 
to cdsarc.u-strasbg.fr (130.79.128.5) or via 
http://cdsweb.u-strasbg.fr/Abastract.html.}}
\subtitle{II. Additional data}}
\author{A.\,C. Gon\c{c}alves, M.-P. V\'eron-Cetty and P. V\'eron
	}
\offprints{A.\,C. Gon\c{c}alves, {\small{anabela@obs-hp.fr}} }
\institute{Observatoire de Haute-Provence (CNRS), F-04870 Saint Michel
l'Observatoire, France}
\date{Received~11 August 1998~/~Accepted 17 November 1998}
\maketitle


\begin{abstract}   

In a previous paper (V\'eron et al. 1997) we presented medium resolution 
(3.4 \AA\ FWHM) spectroscopic observations of 15 ``transition objects'', 
selected for having an ambiguous location in the 
Veilleux \& Osterbrock 
(1987) diagnostic diagrams, and showed that most of them were in fact 
``composite'', this being due to the simultaneous presence on the slit 
of both a Seyfert or Liner nucleus and a \hii\ region. Here, we report new 
spectroscopic observations of 53 emission-line galaxies with a ``transition'' 
spectrum, bringing up to 61 the total number of observed objects in an 
unbiased sample of 88 ``transition objects''. 
Almost all of the observed galaxies have a ``composite" 
nature, confirming the finding that true ``transition'' spectra may not 
exist at all. 

By eliminating ``composite objects'' from the diagnostic diagrams, a clear 
separation between the different classes of nuclear emission-line regions 
(Seyfert 2s, Liners and \hii\ regions) becomes apparent; by restricting the 
volume occupied by the different line-emitting regions in the 3-dimensional 
diagnostic diagrams, we are also restricting the range of possible physical 
parameters in these regions. There seems to be no continuity between 
Seyfert 2s and Liners, the two classes occupying distinct volumes in 
the 3-dimensional space defined by \roiii, \rnii, and \roi.\\


\end{abstract}


\section{Introduction}

The use of the Baldwin et al. (1981) or Veilleux \& Osterbrock (1987) diagnostic
diagrams generally yields an immediate classification of the nuclear 
emission-line clouds; ``transition objects'' exist however, which cannot be 
classified unambiguously from their line ratios (Heckman et al. 1983; Keel 
1984; Veilleux \& Osterbrock 1987; Ho et al. 1993a).
When observed with sufficient spectral resolution, such objects show
diffe\-rent profiles for the permitted and forbidden lines (Heckman et al.
1981; Busko \& Steiner 1990; V\'eron et al. 1981a,b; V\'eron-Cetty \& 
V\'eron 1985, 1986b).

In a previous paper (V\'eron et al. 1997, hereafter Paper I), we 
presented high-dispersion 
(66 \Am) spectroscopic observations of 15 ``transition objects'' 
selected for having an ambiguous location on the Veilleux \& Osterbrock (1987) 
diagnostic diagrams, and showed that most of them are in fact ``composite". 
This was done by modelling the \ha, \nii$\lambda\lambda$\,6548, 6583 and/or 
\hb, \oiii$\lambda\lambda$\,4959, 5007 emission lines with 
Gaussian profiles, allowing for the contribution of several components; best 
fits showed these components to have different line stren\-gths and widths, 
as the result of the lines being produced in regions that are kinematically 
and spatially distinct, usually a Seyfert 2 or Liner cloud and a \hii\ region. 

We have found in the literature 88 emission-line galaxies located north of 
$\delta \sim$ $-$20\degr, with $z <$ 0.100 and $B <$ 17.0, for which 
the published line ratios gave indication of a ``transition'' spectrum, 
constituting an unbiased sample of such objects. Here we report results for 
53 of these galaxies, including seven already observed in Paper I, 
bringing up to 61 the total number of observed objects, or 70\% of 
our unbiased sample of galaxies with a ``transition'' spectrum.


\begin{table*}
\begin{center}
\caption{\label{x_names}
Cross-reference names for the galaxies studied in this paper.}
\begin{flushleft}
\begin{tabular}{crcrrllr}
\hline
NGC & UGC & Zw & MCG\verb+  + & Mark & \verb+  +KUG & 
\verb+     +IRAS & Misc.\verb+ + \\ 
\hline
0034            &  --\verb+  + &   --     & $-$02.01.032  &  938       & 
 \verb+   +--   & \verb+  +00085$-$1223   & --\verb+   + \\

	 --     &  --\verb+  + &   --     & --\verb+   +  &  957       & 
 \verb+   +--   & \verb+  +00391$+$4004   & 5C 3.100     \\

	 --     &  --\verb+  + &   --     & --\verb+   +  & --\verb+ + & 
 \verb+   +--   & \verb+  +01346$-$0924   & --\verb+   + \\

	 --     &   2456$\;$   & 524.040  &   06.07.027   &  1066      &    
 \verb+   +--   & \verb+  +02568$+$3637   & --\verb+   + \\

	 --     &  --\verb+  + &   --     & --\verb+   +  & --\verb+ + &    
 \verb+   +--   & \verb+  +03355$+$0104   & HS 0335$+$0104 \\

	 --     &  --\verb+  + &   --     & --\verb+   +  & --\verb+ + &    
 \verb+   +--   & \verb+  +04210$+$0400   & --\verb+   + \\

	 --     &  --\verb+  + & 420.015  & --\verb+   +  & --\verb+ + &    
 \verb+   +--   & \verb+  +04507$+$0358   & --\verb+   + \\

	 --     &  --\verb+  + &   --     & --\verb+   +  & --\verb+ + &    
 \verb+   +--   & \verb+  +06256$+$6342   & VII Zw 73    \\

	 --     &  4229$\;$    & 207.033  & --\verb+   +  &   622      & 
  0804$+$391    & \verb+  +08043$+$3908   & --\verb+   + \\

	 --     &  --\verb+  + &   --     & --\verb+   +  & --\verb+ + &    
 \verb+   +--   &    \verb+      +$\,-$   & 3C 198.0   \\

	 --     &  --\verb+  + &   --     & --\verb+   +  & --\verb+ + & 
  0825$+$248    &    \verb+      +$\,-$   & --\verb+   + \\

	 --     &  --\verb+  + &   --     & --\verb+   +  & --\verb+ + &    
 \verb+   +--   & \verb+  +09111$-$1007   & --\verb+   + \\

	 --     &  --\verb+  + & 238.066  & --\verb+   +  & --\verb+ + &    
 \verb+   +--   & \verb+  +09277$+$4917   & SBS 0927$+$493 \\

	 --     &  5101$\;$    & 289.011  &   10.14.025   & --\verb+ + &    
 \verb+   +--   & \verb+  +09320$+$6134   & --\verb+   + \\

2989            &  --\verb+  + &   --     & $-$03.25.020  & --\verb+ + &    
 \verb+   +--   & \verb+  +09430$-$1808   & ESO 566$-$G09 \\
    
	 --     &  --\verb+  + &   --     & --\verb+   +  & --\verb+ + &    
 \verb+   +--   & \verb+  +09581$+$3126   & CG 49         \\

3185            &  5554$\;$    & 123.034  &   04.24.024   & --\verb+ + &    
 \verb+   +--   & \verb+  +10148$+$2156   & --\verb+   + \\

\verb+ +  --    &  5984$\;$    & 155.031  &   05.26.024   & --\verb+ + &    
 \verb+   +--   &    \verb+      +$\,-$   & Arp 107A \\

3504            &  6118$\;$    & 155.049  &   05.26.039   & --\verb+ + &  
  1100$+$282    & \verb+  +11004$+$2814   & TEX 1100$+$282 \\

	 --     &  --\verb+  + &   --     & --\verb+   +  & --\verb+ + &    
 \verb+   +--   & \verb+  +11058$-$1131   & --\verb+   + \\

3642            &  6385$\;$    & 291.062  &   10.16.128   & --\verb+ + &    
 \verb+   +--   & \verb+  +11194$+$5920   & --\verb+   + \\

3660            &  --\verb+  + &   --     & $-$01.29.016  &  1291      &    
 \verb+   +--   & \verb+  +11210$-$0823   & --\verb+   + \\

	 --     &  --\verb+  + & 291.074  &   10.17.004   & --\verb+ + &    
 \verb+   +--   & \verb+  +11258$+$5806   & SBS 1125$+$581 \\

	 --     &  --\verb+  + &   --     & --\verb+   +  & --\verb+ + &    
 \verb+   +--   &\verb+  +11285$+$8240\,A & --\verb+   + \\

3758            &  --\verb+  + & 126.110  &  04.27.073    &   739      &    
 \verb+   +--   & \verb+  +11338$+$2152   & --\verb+   + \\

	 --     &  --\verb+  + &   --     & --\verb+   +  & --\verb+ + &    
 \verb+   +--   &    \verb+      +$\,-$   & SBS 1136$+$594 \\

3994            &  6936$\;$    & 186.074  &  06.26.059    & --\verb+ + &  
  1155$+$325\,A &    \verb+      +$\,-$   &  Arp 313     \\

4102            &  7096$\;$    & 269.036  &  09.20.094    & --\verb+ + &    
 \verb+   +--   & \verb+  +12038$+$5259   & --\verb+   + \\

	 --     &  --\verb+  + &   --     & --\verb+   +  & --\verb+ + &    
 \verb+   +--   &\verb+  +12474$+$4345\,S & --\verb+   + \\

	 --     &  8621$\;$    & 218.030  &  07.28.041    & --\verb+ + &    
 \verb+   +--   & \verb+  +13354$+$3924   & --\verb+   + \\

5256            &  8632$\;$    & 246.021  &  08.25.031    &   266      &    
 \verb+   +--   & \verb+  +13362$+$4832   & I Zw  67     \\

	 --     &  --\verb+  + & 073.074  & --\verb+   +  &  1361      &    
 \verb+   +--   & \verb+  +13446$+$1121   & --\verb+   + \\

	 --     &  8718$\;$    & 190.055  &  06.30.085    &   461      &  
  1345$+$343    &   \verb+      +$\,-$    &  CG 1190     \\

	 --     &  --\verb+  + & 162.010  &  05.33.005    & --\verb+ + &    
 \verb+   +--   &  \verb+      +$\,-$     &  4C 26.42    \\

	 --     &  --\verb+  + &   --     & --\verb+   +  & --\verb+ + &    
 \verb+   +--   & \verb+  +14063$+$4905   &  I Zw  81    \\

	 --     &  --\verb+  + & 273.023  & --\verb+   +  &   477      &  
  1439$+$537    & \verb+  +14390$+$5343   &  I Zw  92    \\

	 --     &  --\verb+  + & 221.050  & --\verb+   +  &   848      &    
 \verb+   +--   & \verb+  +15163$+$4255   &  I Zw 107    \\

	 --     &  --\verb+  + & 077.080  & --\verb+   +  & --\verb+ + &    
 \verb+   +--   & \verb+  +15184$+$0834   & --\verb+   + \\

5953            &  9903$\;$    & 107.008  &  03.40.005    &  1512      &    
 \verb+   +--   & \verb+  +15322$+$1521   &  Arp 91     \\

	 --     &  --\verb+  + & 319.034  &  11.19.030    & --\verb+ + &    
 \verb+   +--   & \verb+  +15564$+$6359   &  Kaz  49     \\

	 --     &  --\verb+  + &   --     & --\verb+   +  & --\verb+ + &    
 \verb+   +--   & \verb+  +16129$-$0753   & --\verb+   + \\

	 --     &  --\verb+  + &   --     & --\verb+   +  & --\verb+ + &    
 \verb+   +--   & \verb+  +16382$-$0613   & --\verb+   + \\

	 --     &  10675$\;$   & 169.035  &  05.40.034    &   700      &  
  1701$+$315    & \verb+  +17013$+$3131   & --\verb+   + \\

	 --     &  --\verb+  + & 112.010  &  03.45.003    & --\verb+ + &    
 \verb+   +--   & \verb+  +17334$+$2049   & --\verb+   + \\

	 --     &  --\verb+  + & 142.019  & --\verb+   +  & --\verb+ + &    
 \verb+   +--   & \verb+  +18101$+$2152   & PGC 61548 \\

	 --     &  --\verb+  + & 341.006  & --\verb+   +  & --\verb+ + &    
 \verb+   +--   & \verb+  +18462$+$7207   &  Kaz 214     \\

6764            &  11407$\;$   & 256.007  & 08.35.003     & --\verb+ + &    
 \verb+   +--   & \verb+  +19070$+$5051   & --\verb+   + \\

	 --     &  --\verb+  + &   --     & --\verb+   +  & --\verb+ + &    
 \verb+   +--   & \verb+  +22114$-$1109   & --\verb+   + \\

	 --     &  --\verb+  + & 452.043  &  03.57.031    &   308      &  
  2239$+$199    & \verb+  +22395$+$2000   & --\verb+   + \\ 

	 --     &  --\verb+  + &   --     & --\verb+   +  &   522      &  
  2257$+$161    &  \verb+      +$\,-$     & --\verb+   + \\

7465            &  12317$\;$   & 453.050  &  03.58.024    &   313      &  
  2259$+$156    & \verb+  +22595$+$1541   &  PG 2259$+$156 \\

	 --     &  --\verb+  + & 453.062  & --\verb+   +  & --\verb+ + &    
 \verb+   +--   & \verb+  +23024$+$1916   & --\verb+   + \\

	 --     &  --\verb+  + & 475.036  &  04.54.038    & --\verb+ + &    
 \verb+   +--   & \verb+  +23135$+$2516   &  IC 5298        \\
\hline
\end{tabular}
\end{flushleft}
\end{center}
\end{table*}

\begin{table*}
\begin{center}
\caption{\label{line_ratios}
Published line intensity ratios for the 53 emission-line galaxies 
studied in this paper. References: (1) Aguero et al. (1995), (2) Anton 
(1993), (3) Augarde et al. (1994), (4) Boller et al. (1994), (5) Cohen \& 
Osterbrock  (1981), (6) de Grijp et al. (1992), (7) Delgado \& Perez (1996), 
(8) Duc et al. (1997), (9) Fruscione \& Griffiths (1991), (10) Goodrich \& 
Osterbrock (1983), (11) Hill et al. (1988), (12) Ho et al. (1997a), 
(13) Keel et al. (1985), 
(14) Kim et al. (1995), (15) Klaas \& Elsasser (1991), (16) 
Kollatschny et al. (1983), (17) Koratkar et al. (1995), (18) Koski (1978), 
(19) Martel \& Osterbrock (1994), (20) Netzer et al. (1987), 
(21) Osterbrock \& Pogge (1987), (22) Osterbrock \& Martel (1993), 
(23) Phillips et al. (1983), (24) Rafanelli et al. (1990), 
(25) Salzer et al. (1995), 
(26) Shuder \& Osterbrock (1981), 
(27) Ulvestad \& Wilson (1983), 
(28) Veilleux \& Osterbrock (1987), 
(29) V\'eron et al. (1997), (30) Vogel et al. (1993).}
\begin{flushleft}
\begin{tabular}{lllcc|lllcc}
\hline
Name & \underline{$\lambda$5007$\:$} \verb+ + 
& \underline{$\lambda$6300$\:$} \verb+ +
& \underline{$\lambda$6583$\:$} & Ref. & 
Name & \underline{$\lambda$5007$\:$} \verb+ + 
& \underline{$\lambda$6300$\:$} \verb+ +
& \underline{$\lambda$6583$\:$} & Ref. \\
 & \verb+ +\hb & \verb+ +\ha & \ha & & 
 & \verb+ +\hb & \verb+ +\ha & \ha & \\
\hline
Mark 938             & \verb+ +3.98   & 0.09        & 1.29 	  & (14) & 
IRAS 12474$+$4345\,S & \verb+ +2.93   & \verb+ +--  & 0.42 	  & (6)  \\

Mark 957             & \verb+ +0.63   & 0.04        & 0.46   	  & (19) & 
UGC 8621             & \verb+ +5.25   & 0.081       & 0.95 	  & (22) \\

IRAS 01346$-$0924    & \verb+ +2.90   & \verb+ +--  & 0.41 	  & (6)  & 
Mark  266\,NE        & \verb+ +1.41   & 0.15        & 0.66 	  & (22) \\

Mark 1066            & \verb+ +4.35   & 0.084       & 0.88 	  & (10) &
Mark  266\,SW        & \verb+ +3.98   & 0.05        & 0.50   	  & (22) \\

IRAS 03355$+$0104    &        13.52   & 0.06        & 0.18 	  & (30) & 
Mark 1361            & \verb+ +4.93   & 0.038       & 0.33 	  & (14) \\

\verb+   +"  
&        12.41   & \verb+ +--  & 0.58 	  & (6)  & 
Mark 461             & \verb+  +--    & \verb+ +--  & --	  &  --  \\

IRAS 04210$+$0400    &        14.2    & 0.13        & 0.35	  & (11) &
4C 26.42             & \verb+ +0.40   & 0.22        & 0.81        & (2)  \\

IRAS 04507$+$0358    &        11.77   & \verb+ +--  & 0.28 	  & (6)  &
I Zw 81              & \verb+ +3.14   & 0.066       & 0.67        & (18) \\

VII Zw 73            & \verb+ +3.96   & \verb+ +--  & 0.56 	  & (6)  & 
Mark 477             &        10.42   & 0.17        & 0.36        & (26) \\

Mark~622             & \verb+ +6.25   & 0.064       & 0.94 	  & (26) & 
Mark 848\,S            & \verb+ +1.39   & 0.070       & 0.78        & (14) \\

3C 198.0             & \verb+ +2.13   & 0.04        & 0.35 	  & (5)  & 
IRAS 15184$+$0834    & \verb+ +5.60   & \verb+ +--  & 0.42 	  & (6)  \\

KUG 0825$+$248       & \verb+ +0.84   & 0.13        & 0.28 	  & (3)  & 
NGC 5953             & \verb+ +3.04   & 0.10        & 1.24        & (13) \\

IRAS 09111$-$1007\,E & \verb+ +3.94   & 0.07        & 0.74 	  & (8)  & 
\verb+   +"   
& \verb+ +2.08   & 0.058       & 0.78        & (14) \\

Zw 238.066           & \verb+ +1.73   & 0.065       & 0.82 	  & (14) & 
\verb+   +"   
& \verb+ +4.3    & 0.11        & 1.38        & (7) \\

UGC 5101             & \verb+ +2.86   & 0.089       & 1.34 	  & (14) & 
\verb+   +"   
& \verb+ +4.98   & 0.10        & 1.12        & (24) \\

NGC 2989             & \verb+ +2.50   & 0.037       & 0.52 	  & (23) & 
Kaz 49               & \verb+ +2.58   & 0.025       & 0.56        & (4)  \\

CG 49                &        11.68   & \verb+ +--  & 0.30 	  & (25) & 
IRAS 16129$-$0753    & \verb+ +2.03   & \verb+ +--  & 0.64        & (6)  \\

NGC 3185             & \verb+ +3.42   & 0.045       & 0.70 	  & (12) & 
IRAS 16382$-$0613    & \verb+ +6.67   & 0.09        & 1.09        & (1)  \\

Arp 107A             &        13.62   & 0.38        & 3.00 	  & (13) & 
Mark 700             & \verb+ +0.55   & 0.11        & 1.75        & (18) \\

NGC 3504             & \verb+ +0.53   & 0.023       & 0.59 	  & (12) & 
MCG 03.45.003        & \verb+ +9.74   & \verb+ +--  & 0.42        & (6)  \\

IRAS 11058$-$1131    & \verb+ +9.10   & 0.05        & 0.38 	  & (29) &  
PGC 61548            & \verb+ +1.44   & 0.11        & 0.69        & (9)  \\

NGC 3642             & \verb+ +1.32   & 0.08        & 0.71 	  & (12) & 
Kaz 214              & \verb+ +5.23   & \verb+ +--  & 0.39        & (6)  \\

Mark 1291            & \verb+ +3.18   & \verb+ +--  & 0.48 	  & (16) & 
NGC 6764             & \verb+ +0.53   & 0.045       & 0.68        & (18) \\

IRAS 11285$+$8240\,A & \verb+ +8.25   & 0.106       & 0.46 	  & (15) & 
IRAS 22114$-$1109    & \verb+ +4.22   & 0.077       & 0.62 	  & (14) \\ 
 
Mark 739\,W          & \verb+ +1.14   & \verb+ +--  & 0.43 	  & (20) & 
Mark 308             & \verb+ +4.8    & 0.05        & 0.40        & (29) \\

\verb+   +"  
& \verb+ +1.18   & \verb+ +--  & 0.69 	  & (27) & 
Mark 522             & \verb+ +3.23   & 0.068       & 0.93        & (29) \\

SBS 1136$+$594       &        12.30   & 0.11        & 0.25 	  & (19) & 
Mark 313             & \verb+ +3.52   & 0.10        & 0.52        & (21) \\

NGC 3994             & \verb+ +0.56   & 0.19        & 0.89 	  & (13) & 
Zw 453.062           & \verb+ +1.23   & 0.12        & 1.23        & (14) \\

NGC 4102             & \verb+ +0.99   & 0.041       & 0.92 	  & (12) & 
IC 5298              & \verb+ +4.68   & 0.05        & 0.95        & (14) \\

\hline
\end{tabular}
\end{flushleft}
\end{center}
\end{table*}

\section{Observations and data analysis}

\subsection{Observations}

The 53 observed galaxies suspected to have 
a ``transition'' spectrum are listed in 
Table \ref{x_names} with the various names under which they are 
known, and in Table \ref{line_ratios} with the published line intensity ratios 
\roiii, \roi\ and \rnii. Table \ref{opt_pos} gives their optical positions 
measured on the Digitized Sky Survey\footnotemark[1] 
(V\'eron-Cetty \& V\'eron 1996). 

\footnotetext[1]{The Digitized Sky Survey was produced at the Space 
Telescope Science Institute (STScI) under U.S. Government grant NAG W-2166.}

\begin{table*}[t]
\begin{center}
\caption{\label{opt_pos} 
B1950 optical positions of the observed objects measured on the 
Digitized Sky Survey. The r.m.s. error is 0\farcs6 in each coordinate; 
`` * '' indicates objects with larger errors due to their location near 
one edge of the Schmidt plate (V\'eron-Cetty \& V\'eron 1996). References 
for finding charts: 
(1) Andreasian \& Alloin (1994), (2) Arp (1966), (3) Bowen et al. (1994), 
(4) Carballo et al. (1992), (5) de Grijp et al. (1987),
(6) Delgado \& Perez (1996), (7) Duc et al. (1997), (8) Kazarian (1979),
(9) Keel (1996), (10) Markarian \& Lipovetski
(1971), (11) Markarian \& Lipovetski (1973), (12) Markarian \& Lipovetski
(1974), (13) Markarian et al. (1979a),
(14) Markarian et al. (1979b), (15) Markarian \& Stepanian (1983),
(16) Mazzarella \& Boroson (1993), (17) Olsen (1970), 
(18) Pesch \& Sanduleak (1983), (19) Rubin et al. (1975),
(20) Sandage \& Bedke (1994), (21) Takase \& Miyauchi-Isobe (1986),
(22) Takase \& Miyauchi-Isobe (1990), 
(23) Vogel et al. (1993), (24) Wyndham (1966).}
\begin{flushleft}
\begin{tabular}{lcrcc|lcrcc}
\hline
Name & $\alpha$ & $\delta\; \; \; \; \; \; \;$ & Ref. & mag. & 
Name & $\alpha$ & $\delta\; \; \; \; \; \; \;$ & Ref. & mag. \\
\hline
Mark 938         & 00 08 33.41 & $\/-\,$12 23 06.6 & (16) & 13.5 &
IRAS 12474$+$4345\,S& 12 47 25.08 & $\; $ 43 45 16.6 & (5)  & 15.4 \\

Mark 957         & 00 39 09.65 & $\; $  40 04 51.6 & (3)  & 15.1 &
UGC 8621         & 13 35 28.44 & $\; $  39 24 30.8 &  --  & 14.2 \\

IRAS 01346$-$0924& 01 34 37.53 & $\/-\,$09 24 12.9 & (5)  & 15.8 & 
Mark 266\,SW     & 13 36 14.50 & $\; $  48 31 47.4 & (16) & 13.4 \\

Mark 1066        & 02 56 49.91 & $\; $  36 37 21.1 & (16) & 14.0 & 
Mark 266\,NE     & 13 36 14.99 & $\; $  48 31 53.5 & (16) & 13.4 \\

IRAS 03355$+$0104& 03 35 35.77 & $\; $  01 04 34.0 & (23) & 14.5 &   
Mark 1361        & 13 44 36.53 & $\; $  11 21 20.1 & (14) & 15.3 \\

IRAS 04210$+$0400& 04 21 02.69 & $\; $  04 01 08.2 & (5)  & 16.3 & 
Mark 461         & 13 45 04.29 & $\; $  34 23 51.9 & (9)  & 14.6  \\

IRAS 04507$+$0358& 04 50 47.50 & $\; $  03 58 48.9 & (5)  & 15.0 & 
4C 26.42         & 13 46 33.84 & $\; $  26 50 26.3 & (17) & 15.2 \\

VII Zw 73        & 06 25 37.78 & $\; $  63 42 42.9 & (5)  & 14.8 & 
I Zw 81          & 14 06 20.29 & $\; $  49 05 56.1 &  --  & 16.5 \\

Mark  622        & 08 04 21.03 & $\; $  39 08 57.4 & (12) & 14.1 & 
Mark 477         & 14 39 02.52 & $\; $  53 43 03.3 & (9)  & 15.0 \\

3C 198.0         & 08 19 52.43 & $\; $  06 06 45.7 & (24) & 17.3 &  
Mark 848\,S      & 15 16 19.40 & $\; $  42 55 35.9 & (16) & 15.0 \\

KUG 0825$+$248   & 08 25 29.98 & $\; $  24 48 31.9 & (21) & 16.0 & 
IRAS 15184$+$0834& 15 18 27.10 & $\; $  08 34 33.9 & (5)  & 13.9 \\ 

IRAS 09111$-$1007\,E& 09 11 13.06 & $\/-\,$10 06 54.6 & (7)  & 16.1 &  
NGC 5953         & 15 32 13.23 & $\; $  15 21 35.9 & (6)  & 13.1 \\

Zw 238.066       & 09 27 45.68 & $\; $  49 18 00.4 &  --  & 16.5 & 
Kaz 49           & 15 56 26.70 & $\; $  63 59 00.8 & (8)  & 15.3 \\

UGC 5101         & 09 32 04.95 & $\; $  61 34 36.5 & (9)  & 15.5 &  
IRAS 16129$-$0753& 16 12 58.38 & $\/-\,$07 53 07.2 & (4)  &  --  \\ 

NGC 2989         & 09 43 03.79 & $\/-\,$18 08 35.1 & (20) & 14.5 &  
IRAS 16382$-$0613& 16 38 11.57 & $\/-\,$06 13 07.6 & (4)  & 14.7 \\

CG 49		 & 09 58 07.76 & $\; $  31 26 44.7 & (18) & 16.0 &
Mark 700         & 17 01 21.49 & $\; $  31 31 37.8 & (9)  & 15.5 \\

NGC 3185         & 10 14 53.07 & $\; $  21 56 18.8 & (20) & 12.3 &  
MCG 03.45.003$^{*}$ & 17 33 25.27 & $\; $  20 49 37.6 & (5) & 13.4 \\

Arp 107A         & 10 49 29.66 & $\; $  30 19 25.1 & (2)  & 14.6 &  
PGC 61548        & 18 10 07.06 & $\; $  21 52 15.9 &  --  & 14.2 \\ 

NGC 3504         & 11 00 28.55 & $\; $  28 14 31.6 & (20) & 12.9 &  
Kaz 214          & 18 46 15.77 & $\; $  72 07 42.9 & (5)  & 15.5 \\

IRAS 11058$-$1131& 11 05 49.65 & $\/-\,$11 31 56.8 & (5)  & 14.9 &  
NGC 6764         & 19 07 01.23 & $\; $  50 51 08.5 & (19) & 13.2 \\

NGC 3642         & 11 19 25.03 & $\; $  59 20 54.9 & (20) & 14.1 &  
IRAS 22114$-$1109& 22 11 26.01 & $\/-\,$11 09 21.1 &  --  &  --  \\

Mark 1291        & 11 21 00.13 & $\/-\,$08 23 01.5 & (13) & 15.5 & 
Mark 308         & 22 39 30.53 & $\; $  20 00 00.1 & (10)  & 14.6 \\

IRAS 11285$+$8240\,A & 11 28 41.22 & $\; $  82 40 16.0 &  --  & 15.6 & 
KUG 2239$+$200\,A  & 22 39 33.13 & $\; $  20 00 38.4 & (22) & 15.5 \\

Mark  739\,W$^{*}$ & 11 33 52.49 & $\; $  21 52 22.2 & (16) & 14.0 & 
Mark 522         & 22 57 50.44 & $\; $  16 06 50.7 & (11) & 17.0 \\

SBS 1136$+$594   & 11 36 24.27 & $\; $  59 28 31.4 & (15) & 16.0 &  
Mark 313         & 22 59 32.07 & $\; $  15 41 44.3 & (10) & 14.0 \\

NGC 3994$^{*}$   & 11 55 02.44 & $\; $  32 33 21.1 & (20) & 12.9 &  
Zw 453.062       & 23 02 28.55 & $\; $  19 16 55.2 &  --  & 15.1 \\

NGC 4102         & 12 03 51.33 & $\; $  52 59 21.2 & (20) & 12.6 &  
IC 5298          & 23 13 33.13 & $\; $  25 17 01.9 & (1)  & 14.9 \\
\hline
\end{tabular}
\end{flushleft}
\end{center}
\end{table*}

Spectroscopic observations were carried out during several observing runs 
in May, June and July 1996 and January, March, October and November 1997 
with the spectrograph CA\-RELEC (Lema\^{\i}tre et al. 1989) attached to 
the Cas\-se\-grain focus of the Observatoire de Haute-Pro\-vence (OHP) 1.93 m 
telescope. The detector was a 512 $\times$ 512 pixels, 27 $\times$ 27 $\mu$m  
Tektronic CCD. We used a 600 l\,mm$^{-1}$ grating resulting in a dispersion 
of 66 \Am; the spectral range was 
$\lambda\lambda$\,6700--7600 \AA\ in the red (with a Schott GG 435 filter) 
and $\lambda\lambda$\,4860--5760 \AA\ in the blue. In each case, the galaxy 
nucleus was centered on the slit. 

Usually five columns 
of the CCD ($\sim$ 5\arcsec) were extracted.
The slit width was 2\farcs1, corres\-ponding to a projected slit width on 
the detector of 52 $\mu$m or 1.9 pixel. The slit position angle was not always 
the same for the blue and red spectra; as the aperture used 
is rectangular (2\farcs1 $\times$ 5\arcsec), this may introduce some 
inconsistencies when the line emitting regions are extended. 
The resolution, as measured on the 
night sky emission lines, was $\sim$ 3.4 \AA\ FWHM. The spectra were 
flux calibrated using the standard stars given in Table \ref{std_stars}, 
taken from Oke (1974), Stone  (1977), Oke \& Gunn (1983) and 
Massey et al. (1988). The journal of observations is given in 
Table \ref{journal_obs}.

\subsection{Line profile fitting}

Morgan (1958, 1959) has introduced a classification of galaxies based on 
their nuclear region stellar population. Classes ``a'' and ``af'' 
are dominated by early-type stars. The main absorption features are the 
Balmer lines, which are usually filled up by emission as these objects 
invariably contain a \hii\ region. Classes ``g'', ``gk'' and ``k'' are 
dominated by a bulge of old population II stars. Intermediate classes 
``f'' and ``fg'' have, in addition to a population of young stars, a faint 
bulge of old stars. The old star population have similar spectra in all 
classes (Bica 1988). AGN activity is exceptional in classes ``a'' and 
``af'' but frequent in all other classes (V\'eron \& V\'eron-Cetty 1986); 
in consequence, the nuclear region of most AGNs contains a population of old 
stars with many strong absorption lines which can make the line fitting 
analysis ra\-ther difficult, especially for the 
blue spectra. Therefore, when the absorption blend \mgi\,b $\lambda$5175 
was prominent, we subtracted a suitable fraction of the spectrum of the 
elliptical galaxy NGC 5982, used as a template, to remove the old stellar 
population contribution. Another elliptical, NGC 4365, was used as a template 
for the red spectra, while NGC 821 was used for both the red and blue 
spectra obtained in October and November 1997. All templates were observed 
with the same spectral resolution as the emission-line galaxies. 

Whenever 
ne\-cessary, we have added a \ha\ or \hb\ absorption component; as, 
usually, the \ha\ absorption line is completely filled up by the 
emission lines, we assumed its intensity to be 1.8 times the intensity 
of the nearby absorption \cai\,$\lambda$6495 line (V\'eron-Cetty \& 
V\'eron 1986b). Whenever a template and/or absorption component was used 
in a fit, this is indicated in Table \ref{fits} which contains the line 
fitting analysis results for the 53 observed galaxies. 

\begin{table}[h]
\begin{center}
\caption{\label{std_stars}
Spectrograph settings and standard stars.}
\begin{flushleft}
\begin{tabular}{rcl}
\hline
Date~~~~~~~~~ & $\lambda$ Range (\AA) & \verb+ +Standard stars \\
\hline
21 -- 22.03.95 		& 6500 -- 7400 & \verb+ +BD~26\degr2606 \\
09 -- 10.05.96 		& 6700 -- 7600 & \verb+ +GD~140, BD~26\degr2606\\
11 -- 13.05.96       	& 4860 -- 5760 & \verb+ +Feige~98, Kopff 27\\
      08.06.96          & 4860 -- 5760 & \verb+ +Feige~66, Kopff 27\\
      09.06.96 	        & 6700 -- 7600 & \verb+ +Feige~66, BD~28\degr4211\\
15 -- 16.07.96 		& 4675 -- 5575 & \verb+ +BD~28\degr4211\\
23 -- 25.07.96 		& 6335 -- 7235 & \verb+ +BD~28\degr4211\\
      07.01.97          & 4720 -- 5620 & \verb+ +EG~247 \\
      10.01.97          & 6175 -- 7075 & \verb+ +EG~247 \\
04 -- 07.03.97  	& 4825 -- 5725 & \verb+ +Feige~66 \\
08 -- 12.03.97    	& 6310 -- 7210 & \verb+ +Feige~66 \\
      13.03.97          & 4825 -- 5725 & \verb+ +Feige~66 \\
      31.10.97          & 6455 -- 7365 & \verb+ +Feige~24 \\
01 -- 02.11.97		& 4655 -- 5560 & \verb+ +Feige~24 \\
\hline
\end{tabular}
\end{flushleft}
\end{center}
\end{table}

\begin{table*}
\begin{center}
\caption{\label{journal_obs}
Journal of observations. A: 66 \Am, red; 
B: 66 \Am, blue. A ``*'' indicates objects published in 
Paper I.}
\begin{flushleft}
\begin{tabular}{lcccr|lcccr}
\hline
Name & Disp. & Date & Exp. time & PA &  
Name & Disp. & Date & Exp. time & PA \\
 & & & (min) & (\degr) & & & & (min) & (\degr) \\
\hline
Mark 938	      & A & 31.10.97 & 20 & 90  &
UGC 8621              & A & 10.05.96 & 20 & 180 \\  

                      & B & 01.11.97 & 20 & 180 &
                      & B & 11.05.96 & 20 & 179 \\  

Mark 957              &	A & 10.01.97 & 20 & 90  &
Mark  266\,SW         & A & 10.01.97 & 20 & 180 \\  

                      & B & 01.11.97 & 20 & 270 &
	              & B & 07.03.97 & 20 & 215 \\  

IRAS 01346$-$0924$^{*}$  & A & 10.01.97 & 20 & 90 & 
Mark  266\,NE 	      & B & 07.03.97 & 20 & 215 \\  

Mark 1066             & A & 10.01.97 & 20 & 90  & 
Mark 1361$^{*}$       & B & 06.03.97 & 20 & 180 \\  

                      & B & 07.01 97 & 20 & 90  & 
Mark 461              & A & 22.03.95 & 20 & 90  \\  

IRAS 03355$+$0104        & A & 10.01.97 & 20 & 90  &
                      & B & 08.06.96 & 20 & 89  \\  

	              & B & 04.03.97 & 20 & 90  & 
4C 26.42              & A & 09.06.96 & 20 & 90  \\  

IRAS 04210$+$0400        & A & 10.01.97 & 20 & 90  & 
		      & B & 13.05.96 & 20 & 0   \\  

                      & B & 07.01.97 & 20 & 90  &
I Zw 81               & A & 10.05.96 & 20 & 180 \\  

IRAS 04507$+$0358        & A & 10.01.97 & 20 & 90  & 
		      & B & 11.05.96 & 20 & 84  \\  

                      & B & 06.03.97 & 20 & 180 &
Mark 477$^{*}$        & B & 11.05.96 & 20 & 120 \\  

VII Zw 73             & A & 10.01.97 & 20 & 180 & 
Mark 848\,S           & A & 22.06.96 & 20 & 171 \\  

                      & B & 07.01.97 & 20 & 180 & 
	              & B & 08.06.96 & 20 & 90  \\  

Mark 622              & A & 10.01.97 & 20 & 180 & 
IRAS 15184$+$0834        & A & 08.03.97 & 20 & 270 \\  

                      & B & 07.01.97 & 20 & 180 & 
                      & B & 07.03.97 & 20 & 270 \\  

3C 198.0              & A & 12.03.97 & 20 & 270 &
NGC 5953              & A & 10.05.96 & 20 & 180 \\  

KUG 0825$+$248        & A & 08.03.97 & 20 & 220 & 
                      & B & 11.05.96 & 20 & 180 \\  

                      & B & 07.03.97 & 20 & 217 & 
Kaz 49                & A & 22.06.96 & 20 & 180 \\  

IRAS 09111$-$1007\,E  & A & 08.03.97 & 20 & 258 &
       		      & B & 23.06.96 & 20 & 180 \\  

                      & B & 05.03.97 & 20 & 252 & 
IRAS 16129$-$0753     & A & 22.06.96 & 20 & 180 \\  

Zw 238.066            & A & 08.03.97 & 20 & 180 & 
		      & B & 15.07.96 & 20 & 180 \\  

                      & B & 07.01.97 & 20 & 90  & 
IRAS 16382$-$0613     & A & 09.05.96 & 20 & 180 \\  

UGC 5101              & A & 10.03.97 & 20 & 270 & 
       		      & B & 13.05.96 & 20 & 0  \\   

                      & B & 06.03.97 & 20 & 180 & 
Mark 700              & A & 09.05.96 & 20 & 180 \\  

NGC 2989              & A & 09.03.97 & 20 & 270 &
    		      & B & 07.06.96 & 20 & 90 \\		

                      & B & 06.03.97 & 20 & 180 &
  		      & B & 08.06.96 & 20 & 90 \\

CG 49		      & A & 10.05.96 & 20 & 180 &
MCG 03.45.003$^{*}$   & B & 13.05.96 & 20 & 0   \\		          

		      & B & 01.11.97 & 20 & 270 &
PGC 61548$^{*}$       & A & 09.05.96 & 20 & 180  \\

NGC 3185              & A & 08.03.97 & 20 & 270 & 
		      & B & 13.05.96 & 20 & 0 \\

                      & B & 06.03.97 & 20 & 180 & 
Kaz 214               & A & 09.06.96 & 20 & 139 \\

Arp 107A              & A & 09.03.97 & 20 & 270 & 
		      & B & 13.05.96 & 20 & 0 \\

NGC 3504              & A & 08.03.97 & 20 & 270 & 
NGC 6764              & A & 09.06.96 & 20 & 253 \\

                      & B & 07.03.97 & 20 & 180 & 
		      & B & 08.06.96 & 20 & 72 \\

IRAS 11058$-$1131$^{*}$ & B & 06.03.97 & 20 & 180 & 
IRAS 22114$-$1109       & A & 24.07.96 & 20 & 180 \\

NGC 3642              & A & 08.03.97 & 20 & 270 & 
                      & A & 25.07.96 & 20 & 180 \\

                      & B & 07.03.97 & 20 & 180 & 
		      & B & 15.07.96 & 20 & 180 \\	  

Mark 1291             & A & 10.01.97 & 20 & 180 & 
                      & B & 16.07.96 & 20 & 180 \\	 		      	

                      & B & 06.03.97 & 20 & 180 & 
Mark 308$^{*}$        & A & 09.06.96 & 20 & 44 \\          

IRAS 11285$+$8240\,A  & A & 10.05.96 & 20 & 180 &
Mark 522              & A & 23.07.96 & 20 & 180 \\

Mark  739\,W          & A & 09.01.97 & 20 & 90  & 
                      & B & 15.07.96 & 20 & 180 \\  

	              & B & 07.03.97 & 20 & 272 & 
Mark 313              & A & 23.07.96 & 20 & 180 \\

SBS 1136$+$594        & A & 10.01.97 & 20 & 180 & 
                      & B & 15.07.96 & 20 & 180 \\

                      & B & 05.03.97 & 20 & 180 & 
Zw 453.062	      & A & 25.07.96 & 20 & 225 \\

NGC 3994              & A & 08.03.97 & 20 & 270 & 
	   	      & B & 01.11.97 & 20 & 270 \\

                      & B & 13.03.97 & 20 & 180 & 
	              & B & 02.11.97 & 60 & 270 \\

NGC 4102              & A & 21.03.95 & 20 &  90 &
IC 5298	 	      & A & 25.07.96 & 15 & 204 \\
                      
		      & B & 04.03.97 & 15 & 270 &
                      & B & 02.11.97 & 60 & 270 \\

IRAS 12474$+$4345\,S  & A & 11.03.97 & 20 & 345 &
                      &   &          &    &     \\

                      & B & 13.03.97 & 20 & 345 &
                      &   &          &    &     \\
\hline
\end{tabular}
\end{flushleft}
\end{center}
\end{table*}

\onecolumn
\begin{sidewaystable}
\centering
\small{
\caption{\label{fits}
Fitting profile analysis results. Col. 1 gives the name of the 
object, col. 2 the adopted redshift, cols. 4 and 9 the velocities for each 
set of components measured on the blue and red spectra, respectively, and 
de-redshifted using the redshift given in col.~2; cols. 5 and 10 the 
corresponding FWHM, cols. 6, 11 and 12 the intensity ratios \roiii, 
\rnii\ and \roi\ respectively, and cols. 7 and 13 the 
fraction of the \hb\ emission 
flux (respectively \ha) in each component with respect to the 
total flux of the line in each object. A ``\,T\,'' in col. 3 
(or 8) indicates that the blue (or red) spectrum has been corrected for 
starlight using a suitable fraction of a template (in the blue, we have used 
the elliptical galaxy NGC 5982 and in the red, the elliptical galaxy NGC 4365; 
NGC 821 was used as a template for the objects observed in October and November 
1997); in the same columns, an ``A'' indicates that a \hb\ (or \ha) 
absorption component was added to the model. In col. 14 we give the 
velocity difference between the blue and red systems, and in col. 15 the 
spectroscopic classification of each component in 
the model; Gaussian profiles were 
used throughout, except when indicated by ``$lz$'' (Lorentzian profile). 
Values in parenthesis have been fixed.}
\begin{flushleft}
\begin{tabular}{ll|crrrr|crrccr|rc}
\hline
 Name   &  \verb+  +$z$       &   Stellar     &  $V\; \; \; \; \; \;$         & 
FWHM\/  &   \underline{$\lambda$5007$\:$}     & \hb & Stellar                 & 
$V\; \; \; \; \; \;$          & FWHM\/        & \underline{$\lambda$6583$\:$} &
\underline{$\lambda$6300$\:$} & \ha           & $\Delta V \verb+  + $         &
 Class. \\
        &                     & corr.         & (\kms)                        &
 (\kms) & \hb$\;\;$           & (\%)          & corr.                         &
 (\kms) & (\kms)              & \ha$\;\;$     & \ha$\;\;$                     &
 (\%)                         & (\kms)        &                              \\
\hline
Mark 938        &  0.019  & A  &  111 \verb+  +    &   240 \verb+ +   & 
 1.18\verb+ +   &  91     & T  &  173 \verb+  +    &   270 \verb+ +   & 
 0.71           &  0.07        &   89              & $-$62 \verb+  +  & 
 \hii\           \\

		&  	  &    &($-$79)$\;$ \verb+  +&   890 \verb+ +   &
(10)$\;\verb+  +$ &  9    &    &  $-$11 \verb+  +  &   648 \verb+ +   &
 6.78           & $<$ 0.3\, \verb+  +         & 11 & $-$68 \verb+  +  &
 Sey2	        \\

Mark 957        &  0.0732 &    & $-$402 \verb+  +  &   710 \verb+ +   & 
 9.00\verb+ +   &  8      &    & $-$201 \verb+  +  &   726 \verb+ +   & 
 1.35           & $<$ 0.3\, \verb+  +         & 18 & $-$201 \verb+  + &
 Sey2           \\

		&  	  &    & $-$43  \verb+  +  &   200 \verb+ +   &
 0.15\verb+ +   & 92      &    &   3  \verb+  +    &   192 \verb+ +   &
 0.35           & $<$ 0.02\, \verb+ +         &	82 & $-$46 \verb+  +  &
 \hii\	        \\

IRAS 01346$-$0924&  0.070  &    &  114 \verb+  +    &   178 \verb+ +   & 
 2.13\verb+ +   &  38     &    &  165 \verb+  +    &   148 \verb+ +   & 
 0.70           & $<$ 0.04\, \verb+ +         & 23 & $-$51 \verb+  +  & 
 \hii\           \\

		&  	  &    &  $-$83 \verb+  +  & $<$ 80 \verb+ +  &
 0.83\verb+ +   &  39     &    &  $-$33 \verb+  +  &   219 \verb+ +   &
 0.46           &  0.07        &   37		   & $-$50 \verb+  +  &
 \hii\	        \\

		&  	  &    & $-$224 \verb+  +  &  1069 \verb+ +   &
(10)$\;\verb+  +$ &  23   &    &   -- \verb+   +   &   --  \verb+  +  &
  --            &  --          & --$\;$            &   --  \verb+   + &
 Sey2   	\\

 	        &  	  &    &  -- \verb+   +    &    -- \verb+  +  & 
 -- \verb+  +   & --$\;$  &    &  219 \verb+  +    &  2640 \verb+ +   & 
 --             &   --         &   40              &    -- \verb+   + &
 Sey1		\\

Mark 1066	&  0.0122 &    & $-$116 \verb+  +  &   348 \verb+ +   &
 12.38\verb+ +  &  11     &    &  $-$51 \verb+  +  &   270 \verb+ +   &
 0.77           &  0.07        &   35              & $-$65 \verb+  +  &
 Sey2          \\

		&	  &    &   30 \verb+  +    &   220 \verb+ +   &
 1.54\verb+ +   &  66     &    &   54 \verb+  +    &   205 \verb+ +   &
 0.89           &  0.08   &        39              & $-$24 \verb+  +  &
 \hii\ 	 	\\

		&	  &    & $-$262 \verb+  +  &   875 \verb+ +   &
 5.76\verb+ +   &  23     &    & $-$165 \verb+  +  &  (875)\verb+ +   &
 1.49           &  0.15   &        26              & $-$97 \verb+  +  &
 Sey2	        \\

IRAS 03355$+$0104&  0.0398 &    &   $-$1 \verb+  +  &   364 \verb+ +   &
13.69\verb+ +   &  100    &    &   12 \verb+  +    &   354 \verb+ +   &
 0.49           &  0.12   &        66              & $-$13 \verb+  +  &
 Sey2	 	\\

 	        &  	  &    &  -- \verb+   +    &    -- \verb+  +  & 
 -- \verb+  +   & --$\;$  &    &  264 \verb+  +    &  1930 \verb+ +   & 
 --             &   --    &        34              &    -- \verb+   + &
 Sey1		\\

IRAS 04210$+$0400&  0.0462 &    &   14 \verb+  +    &   178 \verb+ +   &
13.30\verb+ +   &  41     &    &   63 \verb+  +    &   219 \verb+ +   &
 0.40           &  0.17   &        47              & $-$49 \verb+  +  &
 Sey2	 	\\

		&	  &    &    0 \verb+  +    &   541 \verb+ +   &
13.03\verb+ +   &  59     &    &   41 \verb+  +    &   615 \verb+ +   &
 0.34           &  0.06   &        53              & $-$41 \verb+  +  &
 Sey2	 	\\

IRAS 04507$+$0358&  0.0296 &    &  $-$28 \verb+  +  &   240 \verb+ +   &
11.00\verb+ +   &  82     &    &   14 \verb+  +    &   282 \verb+ +   &
 0.47           &  0.09   &        64              & $-$42 \verb+  +  &
 Sey2	        \\

                &         &    & $-$146 \verb+  +  &   603 \verb+ +   & 
14.57\verb+ +   &   18    &    &  -- \verb+   +    &    -- \verb+  +  & 
 --             &   --         & --$\;$            &    -- \verb+   + &
 Sey2	        \\

 	        &  	  &    &  -- \verb+   +    &    -- \verb+  +  & 
 -- \verb+  +   & --$\;$  &    & $-$88 \verb+  +   &  1254$_{lz}$     & 
 --             &   --         &   36		   &    -- \verb+   + &
 Sey1		\\

VII Zw 73	&  0.0405 &    &    0 \verb+  +    &   307 \verb+ +   &
(15)$\;\verb+  +$ &  28   &    &   17 \verb+  +    &   272 \verb+ +   &
 0.80           &  0.10	       &   37              & $-$17 \verb+  +  &
 Sey2	 	\\

		&	  &    & $-$186 \verb+  +  &   794 \verb+ +   &
(15)$\;\verb+  +$ &  12   &    & $-$171 \verb+  +  &   685 \verb+ +   &
 0.52           &  0.22	       &   16              &    15 \verb+  +  &
 Sey2	 	\\

		&	  &    &    3 \verb+  +    &   151 \verb+ +   &
 1.28\verb+ +   &  60     &    &   14 \verb+  +    &   111 \verb+ +   &
 0.48           &  0.02	       &   47              & $-$11 \verb+  +  &
 \hii\	 	\\

Mark 622        &  0.0233 &T, A&    6 \verb+  +    &   178 \verb+ +   & 
(0.1)\,\verb+ + &  59     & T  &   12 \verb+  +    &   205 \verb+ +   & 
 0.79           &  0.06        &   49              &  $-$6 \verb+  +  &
 \hii\	 	\\

		&	  &    &  $-$58 \verb+  +  &  1172 \verb+ +   &
(10)$\;\verb+  +$ &  41   &    &  $-$22 \verb+  +  &   748 \verb+ +   &
 1.44           & $<$ 0.5\, \verb+  +         & 51 & $-$36 \verb+  +  &
 Sey2	 	\\

3C 198.0	&  0.081  &    &  -- \verb+   +    &    -- \verb+  +  &
 -- \verb+  +   & --$\;$  &    &  125 \verb+  +    &   294 \verb+ +   &
 0.28           &  0.05        &  100              &    -- \verb+   + &
 ?	 	\\

KUG 0825$+$248	&  0.083  &    &  108 \verb+  +    &$<$ 80 \verb+ +   &
 0.62\verb+ +   &  100    &    &  135 \verb+  +    &    95 \verb+ +   &
 0.29           &  0.02	       &  100              & $-$27 \verb+  +  &
 \hii\	 	\\

IRAS 09111$-$1007\,E &  0.055 &    & $-$113 \verb+  +  &$<$ 80 \verb+ +   &
(0.1)\,\verb+ + &  16     &    & $-$230 \verb+  +  &    95 \verb+ +   &
 0.55           & $<$ 0.09\, \verb+ +         & 12 &   117 \verb+  +  &
 \hii\	        \\

		&	  &    &($-$76)$\;$ \verb+  +&   259 \verb+ +   &
(0.1)\,\verb+ + &  53     &    &  $-$41 \verb+  +  &   257 \verb+ +   &
 0.86           & $<$ 0.05\, \verb+ + 	      & 40 & $-$35 \verb+  +  &
 \hii\	        \\

		&	  &    & $-$113 \verb+  +  &   430 \verb+ +   &
(10)$\;\verb+  +$ &  31   &    &  $-$96 \verb+  +  &   626 \verb+ +   &
 0.95           &  0.12        &   48		   & $-$17 \verb+  +  &
 Sey2	 	\\

Zw 238.066	&  0.034  &T, A&   95 \verb+  +    &   220 \verb+ +   &
 0.60\verb+ +   &  54     & A  &  149 \verb+  +    &$<$ 80 \verb+ +   &
 0.69           & $<$ 0.05\, \verb+ + 	      & 14 & $-$54 \verb+  +  &
 \hii\	 	\\

		&	  &    & $-$134 \verb+  +  &   154 \verb+ +   &
 1.33\verb+ +   &  37     &    &  $-$71 \verb+  +  &   307 \verb+ +   &
 0.79           &  0.05        &   78              & $-$63 \verb+  +  &
 \hii\	 	\\

		&	  &    & $-$295 \verb+  +  &   935 \verb+ +   &
(10)$\;\verb+  +$ &   9   &    & $-$529 \verb+  +  &  1130 \verb+ +   &
 3.75           & $<$ 0.6\, \verb+  + 	      & 8  &   234 \verb+  +  &
 Sey2	 	\\
\hline
\end{tabular}
\end{flushleft}}
\end{sidewaystable}
\addtocounter{table}{-1}
\newpage
\onecolumn
\begin{sidewaystable}
\centering
\small{
\caption{Fitting profile analysis results (continued).}
\begin{flushleft}
\begin{tabular}{ll|crrrr|crrccr|rc}
\hline
 Name   &  \verb+  +$z$       &   Stellar     &  $V\; \; \; \; \; \;$         & 
FWHM\/  &   \underline{$\lambda$5007$\:$}     & \hb & Stellar                 & 
$V\; \; \; \; \; \;$          & FWHM\/        & \underline{$\lambda$6583$\:$} &
\underline{$\lambda$6300$\:$} & \ha           & $\Delta V \verb+  + $         &
 Class. \\
        &                     & corr.         & (\kms)                        &
 (\kms) & \hb$\;\;$           & (\%)          & corr.                         &
 (\kms) & (\kms)              & \ha$\;\;$     & \ha$\;\;$                     &
 (\%)                         & (\kms)        &                              \\
\hline
UGC 5101 \#1    &  0.039  &    &    -- \verb+   +    &    -- \verb+  +  &
 -- \verb+  +   & --$\;$  &    &    254 \verb+  +    &    95 \verb+ +   &
 0.51           &  0.05        &    100              &    -- \verb+   + & 
 \hii\	        \\

UGC 5101 \#2    &  	  &    &    -- \verb+   +    &    -- \verb+  +  &
 -- \verb+  +   & --$\;$  &    &    222 \verb+  +    &   163 \verb+ +   &
 0.51           &  0.05        &     85              &    -- \verb+   + &
 \hii\ 		\\

		&  	  &    &    -- \verb+   +    &    -- \verb+  +  &
 -- \verb+  +   & --$\;$  &    &    101 \verb+  +    &   526 \verb+ +   &
 2.57           &   $<$ 0.4\, \verb+  +         & 15 &    -- \verb+   + &
 Sey2 		\\

UGC 5101 \#3    &  	  &    &    -- \verb+   +    &    -- \verb+  +  &
 -- \verb+  +   & --$\;$  &    &    146 \verb+  +    &   132 \verb+ +   &
 0.51           &  0.07	       &     28		     &    -- \verb+   + &
 \hii\ 		\\

		&  	  &    &     -- \verb+   +   &    -- \verb+  +  &
 -- \verb+  +   & --$\;$  &    &     92 \verb+  +    &   481 \verb+ +   &
 4.50           &  0.14        &     14              &    -- \verb+   + &
 Sey2 		\\

		&  	  &    &    -- \verb+   +    &    -- \verb+  +  &
 -- \verb+  +   & --$\;$  &    &    197 \verb+  +    &  1033$_{lz}$     &
 --             &   --         &     58  	     &    -- \verb+   + &
 Sey1 	\\

UGC 5101 \#4	&  	  &    &    -- \verb+   +    &    -- \verb+  +  &
 -- \verb+  +   & --$\;$  &    &   $-$108 \verb+  +  &   148 \verb+ +   &
 0.70           & $<$ 0.13\, \verb+ +           & 15 &    -- \verb+   + &
 \hii\ 	        \\

		&  	  &    &    -- \verb+   +    &    -- \verb+  +  &
 -- \verb+  +   & --$\;$  &    &     87 \verb+  +    &   114 \verb+ +   &
 0.38           & $<$ 0.09\, \verb+ +           & 21 &    -- \verb+   + &
	        \\

		&  	  &    &    -- \verb+   +    &    -- \verb+  +  &
 -- \verb+  +   & --$\;$  &    &     36 \verb+  +    &   492 \verb+ +   &
 5.33           & $<$ 0.4\, \verb+  +           & 10 &    -- \verb+   + & 
	 	\\

		&  	  &     &   -- \verb+   +    &    -- \verb+  +  &
 -- \verb+  +   & --$\;$  &     &   324 \verb+  +    &  1438$_{lz}$     &
 --             &   --          &    54              &    -- \verb+   + & 
 Sey1   	\\

UGC 5101 \#5	&  	  &     &   -- \verb+   +    &    -- \verb+  +  &
 -- \verb+  +   & --$\;$  &     &   $-$3 \verb+  +   &   401 \verb+ +   &
 1.31           & $<$ 0.12\, \verb+ +           & 34 &    -- \verb+   + & 
 Sey2 		\\
         	&  	  &     &   -- \verb+   +    &    -- \verb+  +  &
 -- \verb+  +   & --$\;$  &     &  $-$119 \verb+  +  &   148 \verb+ +   &
 0.52           &  0.06         &    66              &    -- \verb+   + &
 \hii\ 		\\

UGC 5101        &  0.039  &  A  &    92 \verb+  +    &   364 \verb+ +   & 
 2.17\verb+ +   &  100    &     &   -- \verb+   +    &    -- \verb+  +  & 
 --             &   --          & --$\;$             &   -- \verb+   +  &
 		\\

NGC 2989	&  0.014  &  A  & $-$105 \verb+  +   &   128 \verb+ +   & 
 1.70\verb+ +   &  100    &     & $-$108 \verb+  +   &   148 \verb+ +   & 
 0.53           &  0.04		&    100             &     3 \verb+  +  &
 \hii\ 		\\

CG 49   	&  0.044  &  A  &  $-$67 \verb+  +   &   259 \verb+ +   & 
 7.20\verb+ +   &  100    &     &  $-$40 \verb+  +   &   257 \verb+ +   & 
 0.80           &  0.09		&    100             & $-$27 \verb+  +  &
 Sey2 		\\

NGC 3185	&  0.004  &  A  &     12 \verb+  +   &   154 \verb+ +   & 
 4.68\verb+ +   &   74    &  A  &     68 \verb+  +   &   148 \verb+ +   & 
 0.81           &  0.04		&     66             & $-$56 \verb+  +  &
  ? 		\\

                &         &     & $-$164 \verb+  +   &   154 \verb+ +   & 
 2.03\verb+ +   &   26    &     &  $-$78 \verb+  +   &   114 \verb+ +   & 
 0.64           & $<$ 0.03\, \verb+ +           & 34 & $-$86 \verb+  +  &
 \hii\ 		\\

Arp 107A	&  0.034  &     &    -- \verb+   +   &    -- \verb+  +  & 
 -- \verb+  +   & --$\;$  &     &   141 \verb+  +    &   178 \verb+ +   & 
 1.41           &  0.20		&    54              &    -- \verb+   + &
 Sey2 		\\

		&  	  &     &    -- \verb+   +   &    -- \verb+  +  & 
 -- \verb+  +   & --$\;$  &     &   108 \verb+  +    &   492 \verb+ +   & 
 1.23           & $<$ 0.1\, \verb+  +           & 46 &    -- \verb+   + &
 Sey2 		\\

NGC 3504	&  0.005  & T, A&   $-$47 \verb+  +  &   200 \verb+ +   & 
(0.1)\,\verb+ + &   95    &  A  &     0 \verb+  +    &   192 \verb+ +   & 
 0.58           &  0.02		&    93              & $-$47 \verb+  +  &
 \hii\ 		\\

		&         &     & $-$10 \verb+  +    &   398 \verb+ +   & 
(10)$\;\verb+  +$ &    5  &     &    12 \verb+  +    &   582 \verb+ +   & 
 1.87           &  0.08		&     7              & $-$22 \verb+  +  &
 Sey2 		\\

IRAS 11058$-$1131&  0.0547 &     &  $-$9 \verb+  +    &   259 \verb+ +   & 
 7.57\verb+ +   &  100    &     &    65 \verb+  +    &   205 \verb+ +   & 
 0.34           &  0.05		&    76              & $-$74 \verb+  +  &
 Sey2 		\\

		&  	  &     &    -- \verb+   +   &    -- \verb+  +  & 
 -- \verb+  +   & --$\;$  &     &   399 \verb+  +    &  2071 \verb+ +   & 
 --             &  --		&    24              &    -- \verb+   + & 
 Sey1		\\

NGC 3642        &  0.005  &  T  &    44 \verb+  +    &    98 \verb+ +   & 
 0.14\verb+ +   &   88    &     &    71 \verb+  +    &    95 \verb+ +   & 
 0.48           &  0.07		&    39              & $-$27 \verb+  +  &
 \hii\ 		\\

	        &         &     &    50 \verb+  +    &  (330)\verb+ +   & 
(10)$\;\verb+  +$ &   12  &     &    36 \verb+  +    &   330 \verb+ +   & 
 1.00           & $<$ 0.3\, \verb+  +           & 9  &    14 \verb+  +  &
 Sey2 		\\

	        &         &     &    -- \verb+   +   &    -- \verb+  +  & 
 -- \verb+  +   & --$\;$  &     & $-$10 \verb+  +    &  2168 \verb+ +   & 
 --             &  --		&    52              &    -- \verb+   + &
 Sey1		\\

Mark 1291       &  0.0122 &     & $-$70 \verb+  +    &   154 \verb+ +   & 
 3.84\verb+ +   &  100    &     &  $-$4 \verb+  +    &   178 \verb+ +   & 
 0.73           &  0.04		&    50              & $-$66 \verb+  +  &
 ?		\\

	        &         &     &    -- \verb+   +   &    -- \verb+  +  & 
 -- \verb+  +   & --$\;$  &     & $-$54 \verb+  +    &  1941 \verb+ +   & 
 --             &  -- 		&    50              &    -- \verb+   + &
 Sey1		\\

IRAS 11285$+$8240\,A &  0.028  &   &    -- \verb+   +   &    -- \verb+  +  & 
 -- \verb+  +   & --$\;$  &     &   141 \verb+  +    &   205 \verb+ +   & 
 0.45           &  0.11		&   100		     &    -- \verb+   + &
 Sey2	 	\\

Mark 739\,W     &  0.0297 &  A  &     9 \verb+  +    &$<$ 80 \verb+ +   & 
 0.27\verb+ +   &   90    &  A  &    14 \verb+  +    &    95 \verb+ +   & 
 0.49           &  0.02		&    44              &   $-$5 \verb+  + &
 \hii\ 		\\

	        & 	  &     &  $-$4 \verb+  +    &   259 \verb+ +   & 
 8.77\verb+ +   &   10    &     & $-$38 \verb+  +    &   307 \verb+ +   & 
 0.67           &  0.10		&    18              &    34 \verb+  +  &
 Sey2 		\\

	        &         &     &    -- \verb+   +   &    -- \verb+  +  & 
 -- \verb+  +   & --$\;$  &     & $-$57 \verb+  +    &  2103 \verb+ +   & 
 --             &  --		&    38              &    -- \verb+   + &
 Sey1		\\

SBS 1136$+$594  &  0.0613 &     & $-$38 \verb+  +    &   154 \verb+ +   & 
11.66\verb+ +   &   10    &     &     6 \verb+  +    &   192 \verb+ +   & 
 0.10           &  0.18		&     9              & $-$44 \verb+  +  &
 Sey2 		\\

	        & 	  &     & $-$105 \verb+  +    &  4470 \verb+ +   & 
 -- \verb+  +   &   90    &     &  $-$21 \verb+  +    &  3871 \verb+ +   & 
 --             &   --		&     83              & $-$84 \verb+  +  &
 Sey1 		\\

	        &         &     &    -- \verb+   +   &    -- \verb+  +  & 
 -- \verb+  +   & --$\;$  &     & $-$23 \verb+  +    &  1010 \verb+ +   & 
 --             &  --		&     8              &    -- \verb+   + &
 Sey1		\\

NGC 3994        &  0.010  &  T  &    98 \verb+  +    &   200 \verb+ +   & 
 0.82\verb+ +   &   89    &  T  &    77 \verb+  +    &   192 \verb+ +   & 
 0.53           &  0.05		&    70              &    21 \verb+  +  &
 \hii\	 	\\

	        &  	  &     & $-$64 \verb+  +    &   905 \verb+ +   & 
 6.00\verb+ +   &   11    &     &    33 \verb+  +    &   537 \verb+ +   & 
 1.41           &  0.53		&    30              & $-$99 \verb+  +  &
 Liner	 	\\

NGC 4102        &  0.0025 &  T  & $-$25 \verb+  +    &   154 \verb+ +   & 
 0.43\verb+ +   &   93    &  A  &    57 \verb+  +    &   192 \verb+ +   & 
 0.87           &  --		&    79              & $-$82 \verb+  +  &
 H II	 	\\

		& 	  &     &$-$126 \verb+  +    &   557 \verb+ +   & 
(10)$\;\verb+  +$ &    7  &     & $-$19 \verb+  +    &  (557)\verb+ +   & 
 1.57           &   --    	&     6              &$-$107 \verb+  +  &
 Sey2	 	\\
 
                &         &     &    -- \verb+   +   &   --  \verb+  +  & 
 -- \verb+  +   & --$\;$  &     &   273 \verb+  +    &   132 \verb+ +   & 
 0.94           &   --		&    15              &   --  \verb+   + &
  ?	 	\\
\hline
\end{tabular}
\end{flushleft}}
\end{sidewaystable}
\addtocounter{table}{-1}
\newpage
\onecolumn
\begin{sidewaystable}
\centering
\small{
\caption{Fitting profile analysis results (continued).}
\begin{flushleft}
\begin{tabular}{ll|crrrr|crrccr|rc}
\hline
 Name   &  \verb+  +$z$       &   Stellar     &  $V\; \; \; \; \; \;$         & 
FWHM\/  &   \underline{$\lambda$5007$\:$}     & \hb & Stellar                 & 
$V\; \; \; \; \; \;$          & FWHM\/        & \underline{$\lambda$6583$\:$} &
\underline{$\lambda$6300$\:$} & \ha           & $\Delta V \verb+  + $         &
 Class. \\
        &                     & corr.         & (\kms)                        &
 (\kms) & \hb$\;\;$           & (\%)          & corr.                         &
 (\kms) & (\kms)              & \ha$\;\;$     & \ha$\;\;$                     &
 (\%)                         & (\kms)        &                              \\
\hline
IRAS 12474$+$4345\,S &  0.062  &     &   182 \verb+  +    &   200 \verb+ +   & 
 2.24\verb+ +   &   53    &     &   128 \verb+  +    &   178 \verb+ +   & 
 0.45           &  0.05		&    47              &    54 \verb+  +  &
 H II	 	\\

		&	  &     &   125 \verb+  +    &   478 \verb+ +   & 
 4.43\verb+ +   &   47    &     &    77 \verb+  +    &   458 \verb+ +   & 
 0.40           &  0.03		&    53              &    48 \verb+  +  &
 ? 		\\

UGC 8621        &  0.020  &  T  &     3 \verb+  +    & $<$ 80 \verb+ +  &
 0.28\verb+ +   &   83    &     &    54 \verb+  +    & $<$ 80 \verb+ +  & 
 0.42           &  $<$ 0.01\, \verb+ +  	& 55 & $-$51 \verb+  +  &
 H II	 	\\

                &         &     & $-$141 \verb+  +   &   664 \verb+ +   & 
(10)$\;\verb+  +$ &   17  &     &  $-$21 \verb+  +   &  (664)\verb+ +   & 
 1.78           &  $<$ 0.3\, \verb+  +  	 & 9 & $-$120 \verb+  + &
 Sey2	 	\\

                &         &     &    -- \verb+   +   &   --  \verb+  +  & 
 -- \verb+  +   & --$\;$  &     & $-$238 \verb+  +   &  2987 \verb+ +   & 
 --             &   -- 	        &    36		     &   --  \verb+   + &
 Sey1		\\

Mark 266 NE     &  0.0283 &     &  $-$88 \verb+  +   &   313 \verb+ +   & 
 0.96\verb+ +   &   63    &     &    -- \verb+   +   &    -- \verb+  +  & 
 --             &   --    	& --$\;$             &    -- \verb+   + &
 H II	 	\\

	        &  	  &     & $-$278 \verb+  +   &  1009 \verb+ +   & 
 2.17\verb+ +   &   37    &     &    -- \verb+   +   &    -- \verb+  +  & 
 --             &   --    	& --$\;$             &    -- \verb+   + &
 Liner? 	\\

Mark 266\,SW    &  0.0278 &     & $-$125 \verb+  +   &   200 \verb+ +   & 
 3.65\verb+ +   &   46    &     &  $-$61 \verb+  +   &   270 \verb+ +   & 
 0.58           &  0.06		&    58              &  $-$64 \verb+  + &
 ?		\\

		&	  &     &   114 \verb+  +    &   295 \verb+ +   & 
 0.36\verb+ +   &   33    &     &   186 \verb+  +    &   270 \verb+ +   & 
 0.58           &  0.04		&    31              &  $-$72 \verb+  + &
 H II	 	\\

		&	  &     & $-$110 \verb+  +   &   603 \verb+ +   & 
13.30\verb+ +   &   21    &     & $-$259 \verb+  +   &  (603)\verb+ +   & 
 0.49           &  0.12		&    11              &   149 \verb+  +  &
 Sey2	 	\\

Mark 1361	& 0.0224  &     &   108 \verb+  +    &   178 \verb+ +   & 
(0.1)\,\verb+ + &   51    &     &   260 \verb+  +    &    95 \verb+ +   & 
 0.54           & 0.05		&    15              & $-$152 \verb+  + &
 H II	 	\\

		& 	  &     &    17 \verb+  +    &   277 \verb+ +   & 
 9.17\verb+ +   &   31    &     &   117 \verb+  +    &   192 \verb+ +   & 
 0.66           &  0.04		&    49              & $-$100 \verb+  + &
 Sey2	 	\\

		& 	  &     & $-$142 \verb+  +   &   785 \verb+ +   & 
(10)$\;\verb+  +$ &   18  &     &    54 \verb+  +    &   548 \verb+ +   & 
 1.04           &  0.09		&    26	             & $-$196 \verb+  + &
 Sey2	 	\\

                &         &     &   -- \verb+   +    &    -- \verb+  +  & 
 -- \verb+  +   & --$\;$  &     &   68 \verb+  +     &  2427 \verb+ +   & 
 --             &   --		&   10               &    -- \verb+   + &
 Sey1		\\

Mark 461        &  0.016  &  T  &   114 \verb+  +    &   313 \verb+ +   & 
 0.99\verb+ +   &   66    &     &    74 \verb+  +    &   219 \verb+ +   & 
 0.60           &   --  	&    37              &    40 \verb+  +  &
 H II	 	\\

                &         &     &  $-$31 \verb+  +   &   920 \verb+ +   & 
$>$ 2.6\verb+   +&   34    &     &    41 \verb+  +   &   692 \verb+ +   & 
 0.64           &   --		&    63              &  $-$72 \verb+  + &
 Sey2?	 	\\

4C 26.42        &  0.063  &  T  & $-$267 \verb+  +   &  (342)\verb+ +   & 
 0.42\verb+ +   &   52    &     & $-$173 \verb+  +   &  (342)\verb+ +   & 
 0.83           &  0.20		&    62              &  $-$94 \verb+  + &
 Liner		\\

                &         &     &    47 \verb+  +    &  (220)\verb+ +   & 
 0.62\verb+ +   &   48    &     &   159 \verb+  +    &  (220)\verb+ +   & 
 0.86           &  0.27   	&    38              & $-$112 \verb+  + &
 Liner		\\

I Zw 81         &  0.052  &  A  &    17 \verb+  +    &  (132)\verb+ +   & 
 2.05\verb+ +   &   53    &     &    98 \verb+  +    &   132 \verb+ +   & 
 0.78           & $<$ 0.04\, \verb+ +   	& 43 &  $-$81 \verb+  + &
 H II?	 	\\

                &         &     & $-$226 \verb+  +   &  (132)\verb+ +   & 
 1.41\verb+ +   &   47    &     & $-$122 \verb+  +   &   132 \verb+ +   & 
 0.56           & $<$ 0.03\, \verb+ +  	        & 57 & $-$104 \verb+  + &
 H II	 	\\

Mark 477        &  0.038  &     & $-$108 \verb+  +   &   200 \verb+ +   & 
 8.00\verb+ +   &   30    &     &  $-$53 \verb+  +   &   163 \verb+ +   & 
 0.38           &  0.15		&    31              &  $-$55 \verb+  + &
 Sey2	 	\\

                &         &     &  $-$83 \verb+  +   &   478 \verb+ +   & 
 11.05\verb+ +  &   47    &     &  $-$26 \verb+  +   &   481 \verb+ +   & 
 0.31           &  0.22		&    47              &  $-$57 \verb+  + &
 Sey2	 	\\

                &         &     & $-$238 \verb+  +   &  1600 \verb+ +   & 
 13.01\verb+ +  &   23    &     & $-$379 \verb+  +   & (1600)\verb+ +   & 
 0.74           &  0.23		&    22		     &   141 \verb+  +  &
 Sey2	 	\\

Mark 848\,S     &  0.040  &     &    81 \verb+  +    &   154 \verb+ +   & 
 0.83\verb+ +   &   93    &     &   128 \verb+  +    &   132 \verb+ +   & 
 0.43           &  0.03		&    81              &  $-$47 \verb+  + &
 H II	 	\\

                &         &     &  $-$92 \verb+  +   &   680 \verb+ +   & 
 4.20\verb+ +   &    7    &     &   122 \verb+  +    &   469 \verb+ +   & 
 0.71           &  0.14		&    19	             & $-$214 \verb+  + &
 Sey2	 	\\

IRAS 15184$+$0834 &  0.031  &  A  &  $-$286 \verb+  +  &  (178)\verb+ +   &
 1.35\verb+ +	&    18   &     &  $-$203 \verb+  +  &   178 \verb+ +   &
 0.54		&  0.01	  	&    21		     &  $-$83 \verb+  + &
 H II		\\  

                &         &     &   $-$6 \verb+  +   &   220 \verb+ +   &
 4.70\verb+ +	&    82   &     &     9 \verb+  +    &   257 \verb+ +   &
 0.69		&  0.05	  	&    79		     &  $-$15 \verb+  + &
  ? 		\\  

NGC 5953        &  0.007  & T, A&  $-$92 \verb+  +   &   200 \verb+ +   & 
 0.55\verb+ +   &   88    &     &  $-$36 \verb+  +   &   205 \verb+ +   & 
 0.60           &  0.03         &    54	  	     &  $-$56 \verb+  + &
 H II	 	\\

                &         &     &  $-$185 \verb+  +  &   398 \verb+ +   & 
(10)$\;\verb+  +$ &   12  &     &  $-$128 \verb+  +  &   412 \verb+ +   & 
 1.96           &  0.18         &    22		     &  $-$57 \verb+  + &
 Sey2	 	\\

                &         &     &   -- \verb+   +    &    -- \verb+  +  & 
 -- \verb+  +   & --$\;$  &     &  311 \verb+  +     &  1768 \verb+ +   & 
 --             &   --		&   24               &    -- \verb+   + &
 Sey1		\\

Kaz 49          &  0.030  &     &    57 \verb+  +    &   259 \verb+ +   & 
 2.21\verb+ +   &   88    &     &   144 \verb+  +    &   232 \verb+ +   & 
 0.55           &  0.05         &    88              &  $-$87 \verb+  + &
 H II	 	\\

                &         &     & $-$252 \verb+  +   &  1069 \verb+ +   & 
(10)$\;\verb+  +$ &   12  &     & $-$218 \verb+  +   &   879 \verb+ +   & 
 2.05           & $<$ 0.5\, \verb+  +           & 12 &  $-$34 \verb+  + &
 Sey2	 	\\

IRAS 16129$-$0753 &  0.033  &     & $-$166 \verb+  +   &   154 \verb+ +   & 
 0.33\verb+ +   &   84    &     &  $-$85 \verb+  +   &   205 \verb+ +   & 
 0.59           & $<$ 0.05\, \verb+ +          & 100 &  $-$81 \verb+  + &
 H II	 	\\

                &         &     & $-$218 \verb+  +   &   649 \verb+ +   & 
 11.25\verb+ +  &   16    &     &    -- \verb+   +   &    -- \verb+  +  & 
 --             &   --    	& --$\;$             &    -- \verb+   + &
 Sey2	 	\\

IRAS 16382$-$0613 &  0.028  &     &  $-$76 \verb+  +   &   348 \verb+ +   & 
 3.94\verb+ +   &   40    &     &  $-$72 \verb+  +   &   257 \verb+ +   & 
 0.91           & $<$ 0.03\, \verb+ +   	& 24 &   $-$6 \verb+  + &
 ?		\\

                &         &     & $-$310 \verb+  +   &  1157 \verb+ +   & 
 4.06\verb+ +   &   60    &     & $-$188 \verb+  +   &  1064 \verb+ +   & 
 1.41           &  $<$ 0.12\, \verb+ +          & 24 & $-$122 \verb+  + &
 Sey2		\\

                &         &     &    -- \verb+   +   &    -- \verb+  +  & 
 -- \verb+  +   & --$\;$  &     &   144 \verb+  +    &  4980 \verb+ +   & 
 --             &   --  	&    52		     &    -- \verb+   + &
 Sey1		\\
\hline
\end{tabular}
\end{flushleft}}
\end{sidewaystable}
\addtocounter{table}{-1}
\newpage
\onecolumn
\begin{sidewaystable}
\centering
\small{
\caption{Fitting profile analysis results (end).}
\begin{flushleft}
\begin{tabular}{ll|crrrr|crrccr|rc}
\hline
 Name   &  \verb+  +$z$       &   Stellar     &  $V\; \; \; \; \; \;$         & 
FWHM\/  &   \underline{$\lambda$5007$\:$}     & \hb & Stellar                 & 
$V\; \; \; \; \; \;$          & FWHM\/        & \underline{$\lambda$6583$\:$} &
\underline{$\lambda$6300$\:$} & \ha           & $\Delta V \verb+  + $         &
 Class. \\
        &                     & corr.         & (\kms)                        &
 (\kms) & \hb$\;\;$           & (\%)          & corr.                         &
 (\kms) & (\kms)              & \ha$\;\;$     & \ha$\;\;$                     &
 (\%)                         & (\kms)        &                              \\
\hline
Mark 700        &  0.034  &  A  &  $-$74 \verb+  +   &   618 \verb+ +   & 
 1.86\verb+ +   &  100    &     &  $-$28 \verb+  +   &   435 \verb+ +   & 
 2.19           &  0.19         &    23		     &  $-$46 \verb+  + &
 Liner		\\

                &         &     &    -- \verb+   +   &    -- \verb+  +  & 
 -- \verb+  +   & --$\;$  &     &   369 \verb+  +    &  1595 \verb+ +   & 
 --             &   --          &    61              &    -- \verb+   + &
 Sey1		\\

                &         &     &    -- \verb+   +   &    -- \verb+  +  & 
 -- \verb+  +   & --$\;$  &     & $-$1137 \verb+  +  &  1097 \verb+ +   & 
 --             &   --          &    16              &    -- \verb+   + &
 Sey1		\\

MCG 03.45.003   &  0.024  &     &    87 \verb+  +    &    98 \verb+ +   & 
 14.01\verb+ +  &   47    &     &    77 \verb+  +    &   132 \verb+ +   & 
 0.51           &  0.17         &    54		     &    10 \verb+  +  &
 Sey2	 	\\

                &         &     &    27 \verb+  +    &   364 \verb+ +   & 
 8.71\verb+ +   &   53    &     &    39 \verb+  +    &   435 \verb+ +   & 
 0.61           & $<$ 0.07\, \verb+ +           & 46 &  $-$12 \verb+  + &
 Sey2	 	\\

PGC 61548       &  0.018  &  T  &    20 \verb+  +    &   277 \verb+ +   & 
 0.41\verb+ +   &   92    &     &    68 \verb+  +    &   282 \verb+ +   & 
 0.50           &  0.04		&    93              &  $-$48 \verb+  + &
 H II	 	\\

                &         &     &   $-$5 \verb+  +   &   541 \verb+ +   & 
(10)$\;\verb+  +$ &    8  &     &    14 \verb+  +    &   604 \verb+ +   & 
 3.94           &  0.49		&     7              &  $-$19 \verb+  + &
 Sey2	 	\\

Kaz 214         & 0.046   &  A  &     0 \verb+  +    &$<$ 80 \verb+ +   & 
 5.32\verb+ +   &   10    &     &    -- \verb+   +   &    -- \verb+  +  & 
 --             &   --    	& --$\;$             &    -- \verb+   + &
 ?		\\

                &         &     &    98 \verb+  +    &   330 \verb+ +   & 
 3.60\verb+ +   &   73    &     &    -- \verb+   +   &    -- \verb+  +  & 
 --             &   --    	& --$\;$             &    -- \verb+   + &
 ?		\\

                &         &     &   195 \verb+  +    &   740 \verb+ +   & 
(10)$\;\verb+  +$ &   17  &     &    -- \verb+   +   &    -- \verb+  +  & 
                &   --    	& --$\;$             &    -- \verb+   + &
 Sey2		\\

Kaz 214 \#1     &         &     &    -- \verb+   +   &    -- \verb+  +  & 
 -- \verb+  +   & --$\;$  &     &  $-$25 \verb+  +   &    95 \verb+ +   & 
 0.32           &  $<$ 0.06\, \verb+ + 	       & 100 &    -- \verb+   + &
 H II		\\

Kaz 214 \#2     &         &     &    -- \verb+   +   &    -- \verb+  +  & 
 -- \verb+  +   & --$\;$  &     &  $-$28 \verb+  +   &   148 \verb+ +   & 
 0.38           &  $<$ 0.05\, \verb+ + 	       & 100 &    -- \verb+   + &
 H II		\\

Kaz 214 \#3     &         &     &    -- \verb+   +   &    -- \verb+  +  & 
 -- \verb+  +   & --$\;$  &     &    12 \verb+  +    &   245 \verb+ +   & 
 0.52           &  $<$ 0.05\, \verb+ + 	        & 68 &    -- \verb+   + &
 H II		\\

                &         &     &    -- \verb+   +   &    -- \verb+  +  & 
 -- \verb+  +   & --$\;$  &     &  (120)\verb+  +    &  (525)\verb+ +   & 
(0.60)          &  0.17		&    32              &    -- \verb+   + &
 Sey2?		\\

Kaz 214 \#4     &         &     &    -- \verb+   +   &    -- \verb+  +  & 
 -- \verb+  +   & --$\;$  &     &    77 \verb+  +    &   205 \verb+ +   & 
 0.33           &  $<$ 0.03\, \verb+ + 	        & 60 &    -- \verb+   + &
 H II		\\

                &         &     &    -- \verb+   +   &    -- \verb+  +  & 
 -- \verb+  +   & --$\;$  &     &  (120)\verb+  +    &  (525)\verb+ +   & 
(0.60)          & 0.16		&    40              &    -- \verb+   + &
 Sey2?		\\

Kaz 214 \#5     &         &     &    -- \verb+   +   &    -- \verb+  +  & 
 -- \verb+  +   & --$\;$  &     &   114 \verb+  +    &   132 \verb+ +   & 
 0.28           & $<$ 0.02\, \verb+ + 	        & 45 &    -- \verb+   + &
 H II		\\

                &         &     &    -- \verb+   +   &    -- \verb+  +  & 
 -- \verb+  +   & --$\;$  &     &  (120)\verb+  +    &  (525)\verb+ +   & 
(0.60)          &  0.09		&    55              &    -- \verb+   + &
 Sey2?		\\

Kaz 214 \#6     &         &     &    -- \verb+   +   &    -- \verb+  +  & 
 -- \verb+  +   & --$\;$  &     &   135 \verb+  +    &   114 \verb+ +   & 
 0.31           &  0.05		&    44              &    -- \verb+   + &
 H II		\\

                &         &     &    -- \verb+   +   &    -- \verb+  +  & 
 -- \verb+  +   & --$\;$  &     &  (120)\verb+  +    &  (525)\verb+ +   & 
(0.60)          &  0.16 	&    56              &    -- \verb+   + &
 Sey2?		\\

Kaz 214 \#7     &         &     &    -- \verb+   +   &    -- \verb+  +  & 
 -- \verb+  +   & --$\;$  &     &   141 \verb+  +    &   232 \verb+ +   & 
 0.35           & $<$ 0.06\, \verb+ + 	       & 100 &    -- \verb+   + &
 H II		\\

NGC 6764        &  0.008  &     &    47 \verb+  +    &   330 \verb+ +   & 
 0.62\verb+ +   &   72    &     &   $-$7 \verb+  +   &   319 \verb+ +   & 
 0.65           &  0.04		&    66              &    54 \verb+  +  &
 H II		\\

		&	  &     & $-$158 \verb+  +   &   430 \verb+ +   & 
 0.44\verb+ +   &   28    &     & $-$101 \verb+  +   &   537 \verb+ +   & 
 0.96           &  0.14		&    34              &  $-$57 \verb+  + &
 Liner		\\

IRAS 22114$-$1109 &  0.054  &     &    44 \verb+  +    &   154 \verb+ +   & 
 1.33\verb+ +   &   46    &     &  $-$16 \verb+  +   &   219 \verb+ +   & 
 0.70           &  $<$ 0.07\, \verb+ + 	        & 57 &    60 \verb+  +  &
 H II		\\

		&	  &     &   195 \verb+  +    &   510 \verb+ +   & 
(10)$\;\verb+  +$ &   54  &     &   203 \verb+  +    &   319 \verb+ +   & 
 0.60           &  0.12		&    43              &   $-$8 \verb+  + &
 Sey2	 	\\

Mark 308        &  0.023  &     &   155 \verb+  +    &   178 \verb+ +   & 
(0.1)\,\verb+ + &   56    &     &    87 \verb+  +    &   132 \verb+ +   & 
 0.30           &  0.06   	&    44              &    68 \verb+  +  &
 H II		\\
 
		&	  &     &   243 \verb+  +    &   240 \verb+ +   & 
(10)$\;\verb+  +$ &   24  &     &    63 \verb+  +    &   412 \verb+ +   & 
 0.49           &   0.03	&    30              &   180 \verb+  +  &
 Sey2?	 	\\

		&	  &     &   182 \verb+  +    &   995 \verb+ +   & 
(10)$\;\verb+  +$ &   20  &     &  $-$34 \verb+  +   &  1097 \verb+ +   & 
 0.50           &  0.19 	&    19              &   216 \verb+  +  &
 Sey2	 	\\
 
                &         &     &    -- \verb+   +   &    -- \verb+  +  & 
 -- \verb+  +   & --$\;$  &     &   (23)\verb+  +    & (1725)\verb+ +   & 
 --             &  -- 		&     7              &    -- \verb+   + &
 Sey1		\\

Mark 522 	&  0.032  &  A  &    95 \verb+  +    &$<$ 80 \verb+ +   &
 0.63\verb+ +   &   86    &     &   114 \verb+  +    &   114 \verb+ +   &
 0.53           &  $<$ 0.04\, \verb+ +  	& 75 &  $-$19 \verb+  + &
 H II		\\

		& 	  &     &  $-$15 \verb+  +   &   220 \verb+ +   & 
 7.87\verb+ +   &   14    &     &    30 \verb+  +    &   270 \verb+ +   & 
 1.50           &  $<$ 0.2\, \verb+  +  	& 25 &  $-$45 \verb+  + &
 Sey2	 	\\

Mark 313        &  0.006  &  A  &   128 \verb+  +    &   154 \verb+ +   & 
 2.29\verb+ +   &  100    &     &   117 \verb+  +    &   114 \verb+ +   & 
 0.44           &  0.10 	&    74              &    11 \verb+  +  &
 H II		\\

		&	  &     &   -- \verb+   +    &    -- \verb+  +  & 
 -- \verb+  +   & --$\;$  &     &    84 \verb+  +    &   424 \verb+ +   & 
 0.71           &  0.28		&    26              &    -- \verb+   + &
 Sey2?		\\

Zw 453.062 	&  0.025  &  A  &    20 \verb+  +    &   200 \verb+ +   &
 0.31\verb+ +   &   94    &     &  $-$42 \verb+  +   &   163 \verb+ +   &
 0.46           &  0.05         &    91              &    62 \verb+  +  &
 H II		\\

		&	  &     & $-$274 \verb+  +   &   541 \verb+ +   & 
(10)$\;\verb+  +$ &    6  &     & $-$160 \verb+  +   &   514 \verb+ +   & 
 3.33           &   0.4 $\;$   	                & 9  &   114 \verb+  +  &
 Sey2		\\

IC 5298 	&  0.027  &  T  &    95 \verb+  +    &   313 \verb+ +   &
 0.53\verb+ +   &   76    &     &    44 \verb+  +    &   307 \verb+ +   &
 0.83           &   0.12        &    86              &    51 \verb+  +  &
 H II  		\\

   	        &         &     &  $-$68 \verb+  +   &   557 \verb+ +   &
(10)$\;\verb+  +$ &   24  &     &    33 \verb+  +    &   537 \verb+ +   &
 2.28           &  $<$ 0.4\, \verb+  +  	& 14 & $-$101 \verb+  + &
 Sey2		\\
\hline
\end{tabular}
\end{flushleft}}
\end{sidewaystable}
\newpage
\twocolumn  

The line fitting analysis of the spectra was done in terms of Gaussian 
components as described in V\'eron et al. (1980, 1981b,c). The three 
emission lines, \ha\ and \nii$\lambda\lambda$\,6548, 6583 (or \hb\ and 
\oiii$\lambda\lambda$\,4959, 5007) 
were fitted by one or several sets of three Gaussian components; 
whenever necessary, two components were added to fit the 
\oi $\lambda\lambda$\,6300, 6363 lines. The width and redshift of each 
component in a set were supposed to be the same. The intensity ratios 
of the \nii$\lambda\lambda$\,6548, 6583, \oiii$\lambda\lambda$\,4959, 5007 and 
\oi$\lambda\lambda$\,6300, 6363 lines were taken to be 
equal to 3.00, 2.96 and 3.11, respectively (Osterbrock 1974). Whenever a fit 
resulted in a small negative intensity for a \hb\ component, we set the 
corresponding \roiii\ ratio to 10, the mean value for Seyfert 2s. 

All line widths given in this paper have been 
corrected for the instrumental broadening. The spectra and best 
fits are plotted in Fig. \ref{spectra}, the parameters describing the 
individual components required by the analysis being given in Table \ref{fits}.

\subsection{Notes on individual objects}

{\bf Mark 938}. This galaxy is apparently undergoing a merger as evidenced by 
the presence of tidal tails (Mulchaey et al. 1996; Mazzarella \& Boroson 1993). 
The nature of its emission-line spectrum has been rather controversial. 
Afa\-nasjev et al. (1980) classified it as a Seyfert 2, Osterbrock \& Dahari 
(1983) claimed that it is not a Seyfert; Dahari (1985), V\'eron-Cetty \& 
V\'eron (1986a) and Veilleux et al. (1995) called it again a Seyfert 2, while 
Mulchaey et al. (1996), observing a weak emission of \oiii$\lambda$5007 and 
a strong 
\ha\ over the entire galaxy, suggested that there is no Seyfert activity 
in this object, in agreement with Mazzarella \& Boroson (1993) who called it 
a \hii\ region. The line ratios published by Veilleux et al. (1995) and Vaceli 
et al. (1997) indicate a ``transition'' spectrum, the \oiii\ lines 
being weak for a Seyfert 2 galaxy (\roiii\ = 4). The high resolution 
spectroscopic observations of Busko \& Steiner (1990), showing complex emission 
line profiles with great differences in width and shape between \ha\ and 
\nii$\lambda$6583 (the measured line widths are 264 $\pm$7 and 
384 $\pm$12 \kms\ for \ha\ and \nii$\lambda$6583, respectively), suggest 
a ``composite'' spectrum. To fit our spectra, two components are needed: one 
is a \hii\ region with narrow lines ($\sim$ 255 \kms\ FWHM); 
the other is a Seyfert 2 with much broader 
lines ($\sim$ 760 \kms\ FWHM). For this component, we find a very high and 
unlikely \rnii\ ratio ($\sim$ 6.8); however, there is a very strong and broad 
\hb\ absorption line. It is probable that the broad \ha\ emission component 
intensity is greatly reduced by the presence of a \ha\ absorption line 
which has not been accounted for.

{\bf Mark 957}. This galaxy has been identified with the radiosource 5C 3.100 
(Antonucci 1985) and a ROSAT X-ray sour\-ce (Boller et al. 1998). 
Dahari \& de Robertis (1988) called it a Seyfert 2. However, Koski (1978) 
and Halpern \& Oke (1987) have observed strong \feii\ emission lines in 
this object; furthermore, the continuum is very flat, extending far into the 
blue (Koski 1978), accounting 
for the classification of this object as a Narrow Line Seyfert 1 galaxy. This 
classification is supported by Boller et al. (1996) who have found a steep 
soft X-ray component (photon index $\Gamma$ = 2.9 $\pm$ 0.2) with a variable 
flux (by a factor 1.9 over 18\,900 sec). The \hb\ line is very narrow 
(FWHM $<$ 685 \kms) (Goodrich 1989); narrow \ha\ and \nii\ lines are observed 
as far as 10\arcsec\ from the nucleus (with \rnii\ $\sim$ 0.4) (Halpern \& 
Oke 1987), suggesting the presence of an extended \hii\ region. In the 
nucleus, the high ionization lines (\oiii\ and \neiii$\lambda$3869) are 
found to be blueshifted by $\sim$ 280 \kms\ with respect to the low 
ionization lines. Although having a relatively low signal-to-noise ratio, 
our spectra are quite interesting. In the blue, there is a very narrow \hb\ 
emission line (FWHM $\sim$ 200 \kms) associated with very weak (\roiii\ 
$\sim$ 0.15) and relatively broad (FWHM $\sim$ 710 \kms) \oiii\ lines; the 
associated broad \hb\ component is 
weak (\roiii\ $\sim$ 9) and accounts for only 8\% of the total \hb\ flux. 
The \oiii\ lines are blueshifted by $\sim$ 360 \kms\ with respect to \hb. 
The red spectrum is also reasonably fitted with two sets of components; one 
is narrow with weak \nii\ lines, while the second is broader with 
relatively strong \nii\ lines. This is in satisfactory agreement with Halpern 
\& Oke's results, and suggests that the nuclear spectrum is dominated by a 
strong \hii\ region superimposed onto a relatively weak Seyfert 2 nucleus.

{\bf IRAS 01346\,-\,0924} was identified by de Grijp 
et al. (1987) with a galaxy they called MCG $-$02.05.022, which se\-ems to be 
erroneous. It was classified a Seyfert 2 by de Grijp et al. (1992) on the 
basis of its emission-line ratios. We discussed this object in Paper I, 
giving it the wrong name (MCG $-$02.05.022); we suggested, on the basis of a 
blue spectrum, that it was ``composite''. The best blue spectrum fit 
is obtained with three sets of three Gaussians, 
two being typical of a \hii\ region and the third of a weak Seyfert 2 
nebulosity. A weak broad (FWHM $\sim$ 2\,640 \kms) \ha\ component may also 
be present. The Seyfert 2 cloud is so weak that it is not detected on our 
red spectrum.

\begin{figure*}
\resizebox{18cm}{!}{\includegraphics{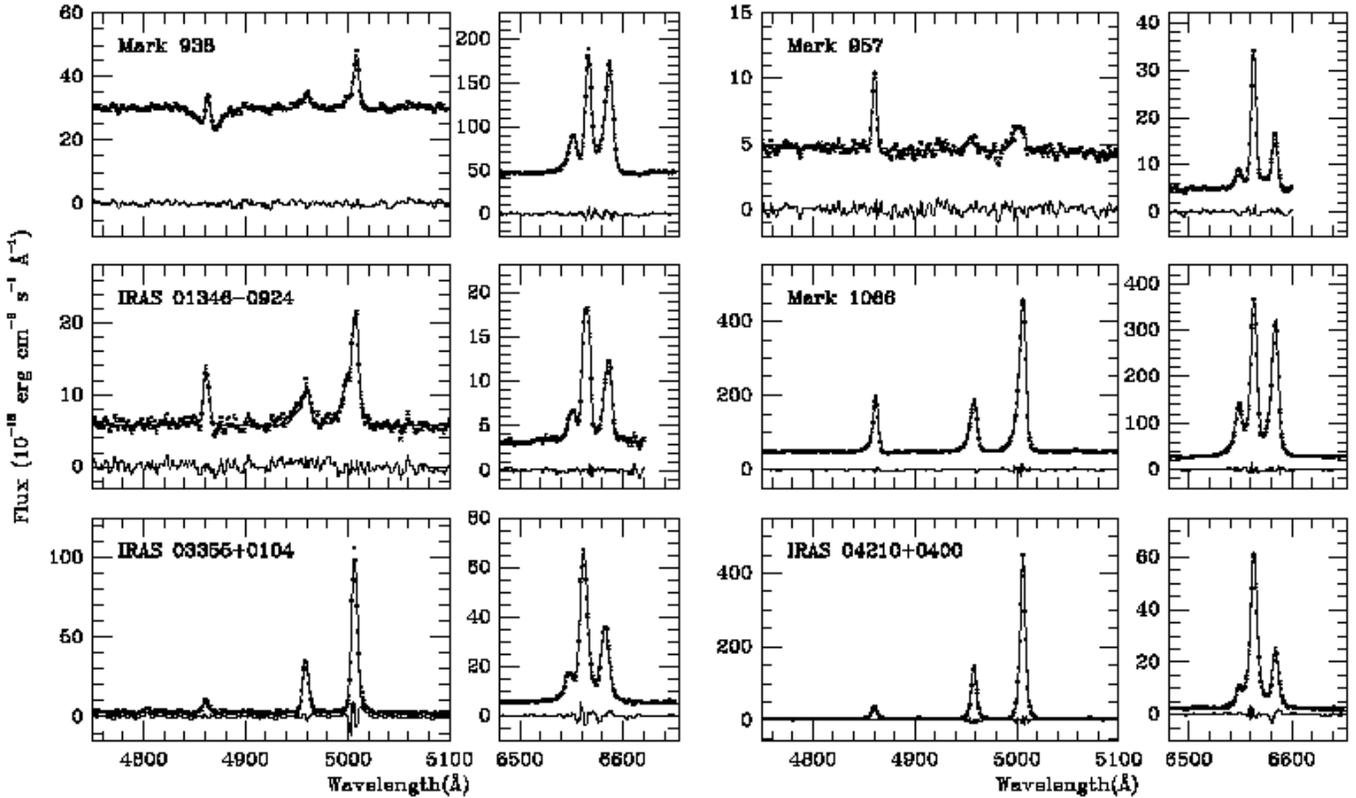}}
\caption{\label{spectra}
Blue and red spectra for the 53 galaxies studied in this paper. 
For 52 of the observed AGNs we present 66 \Am\ spectra; in the case of 
VII Zw 73, we give 33 \Am\ spectra. For 4 of the observed objects, only 
one spectral region is available. All the spectra were de-redshifted to 
rest wavelengths. The spectral ranges displayed are 
$\lambda\lambda$\,4750--5120 \AA\ and $\lambda\lambda$\,6480--6650 \AA. In 
each frame the data points (small crosses), the best fit (solid line) and 
the residuals (lower solid line) are shown. For the red spectrum of 
SBS 1136$+$594, the individual components of the fit are also given as 
an example (dotted lines).}
\end{figure*}
\addtocounter{figure}{-1}
\begin{figure*}
\resizebox{18cm}{!}{\includegraphics{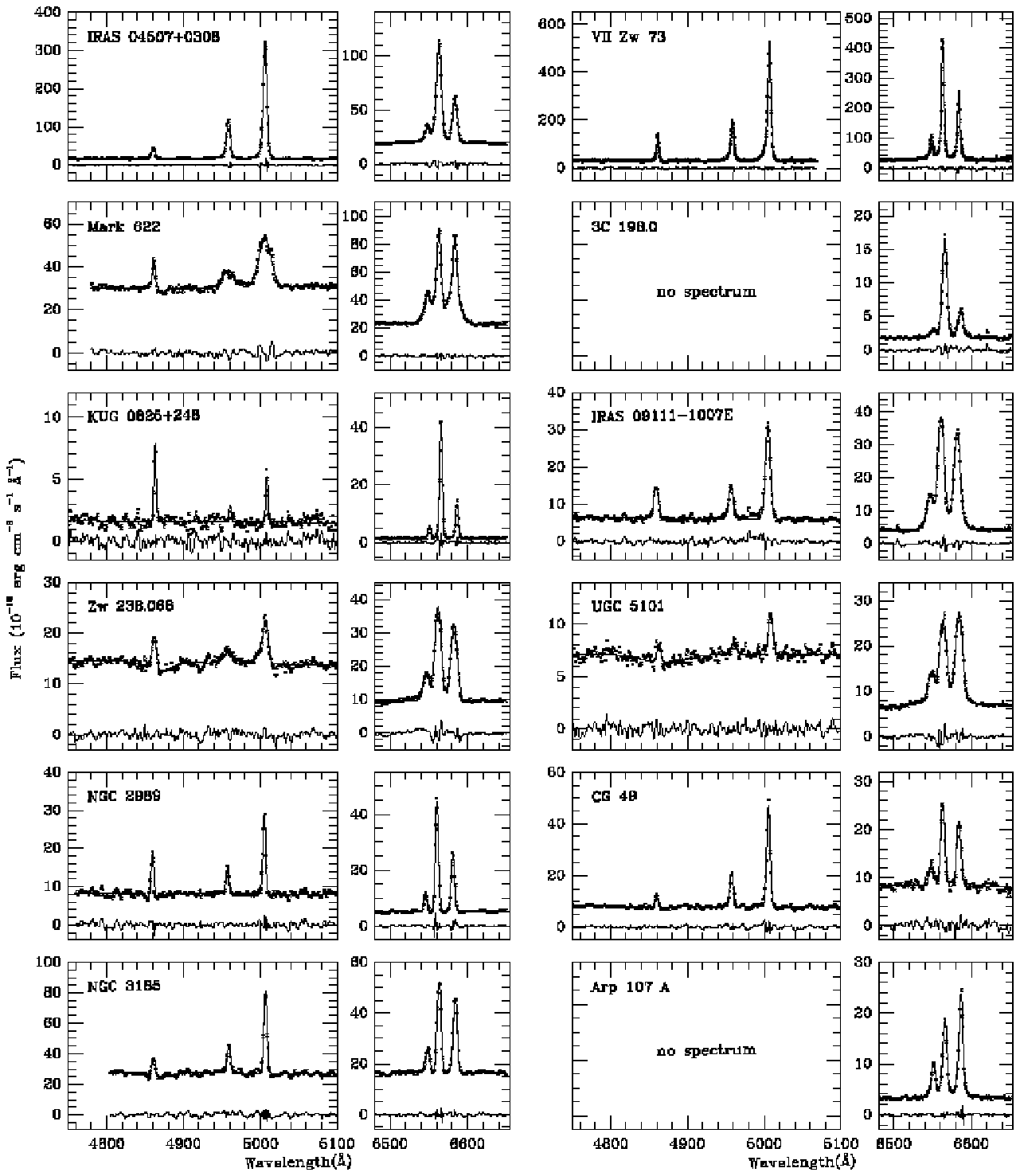}}
\caption{Blue and red spectra for the 53 galaxies studied in this paper 
(continued).}
\end{figure*}
\addtocounter{figure}{-1}
\begin{figure*}
\resizebox{18cm}{!}{\includegraphics{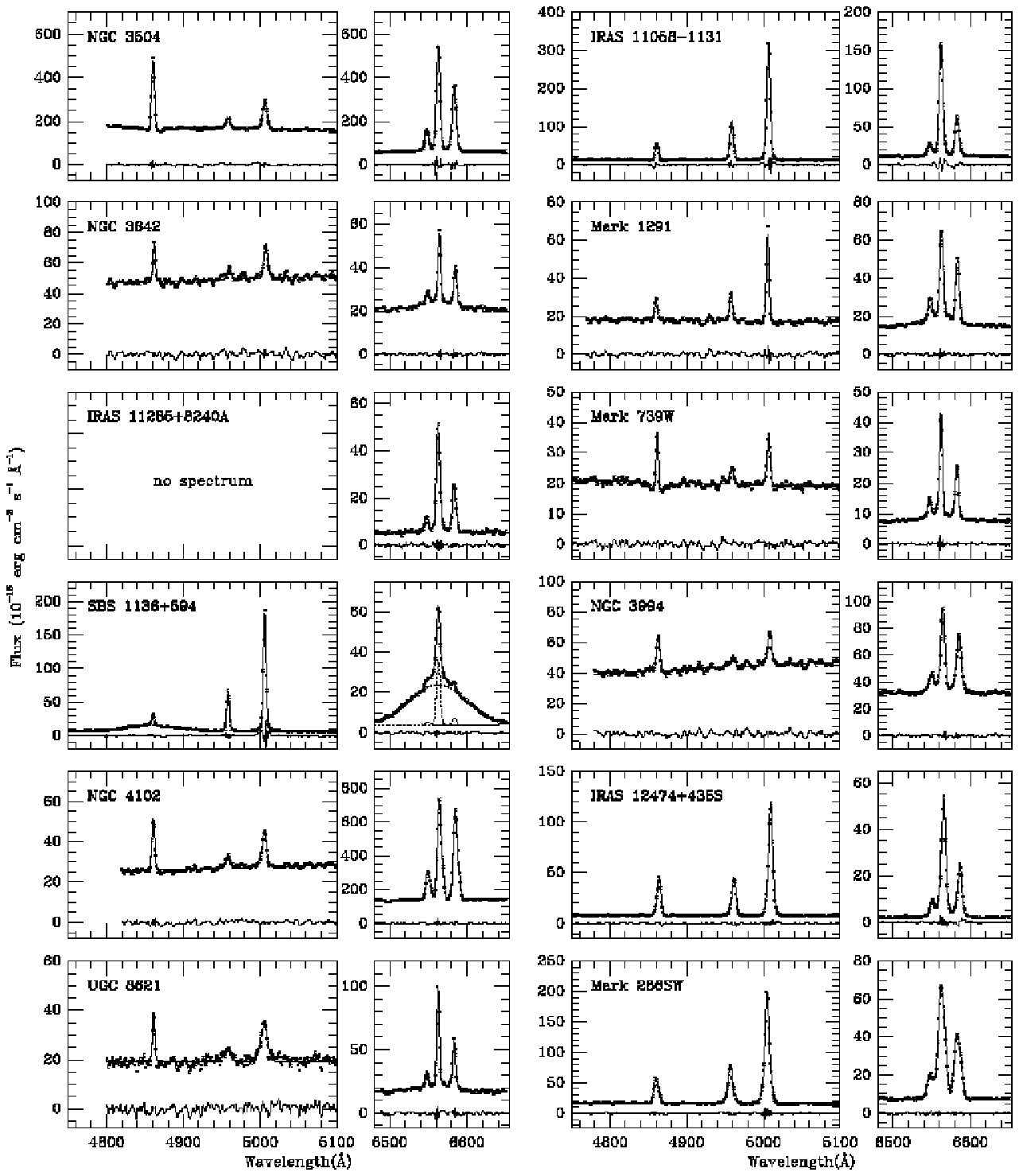}}
\caption{Blue and red spectra for the 53 galaxies studied in this paper 
(continued).}
\end{figure*}
\addtocounter{figure}{-1}
\begin{figure*}
\resizebox{18cm}{!}{\includegraphics{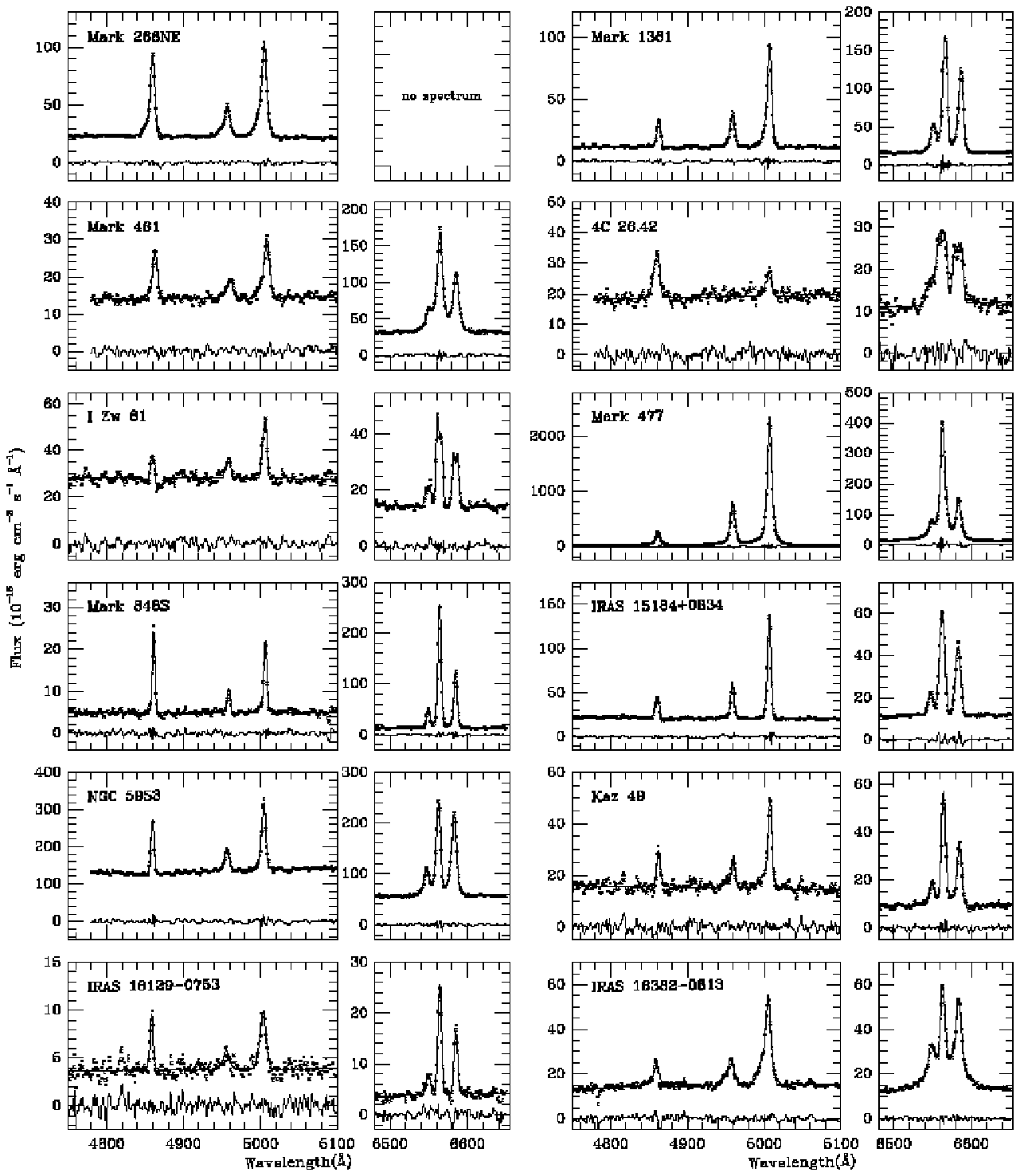}}
\caption{Blue and red spectra for the 53 galaxies studied in this paper 
(continued).}
\end{figure*}
\addtocounter{figure}{-1}
\begin{figure*}
\resizebox{18cm}{!}{\includegraphics{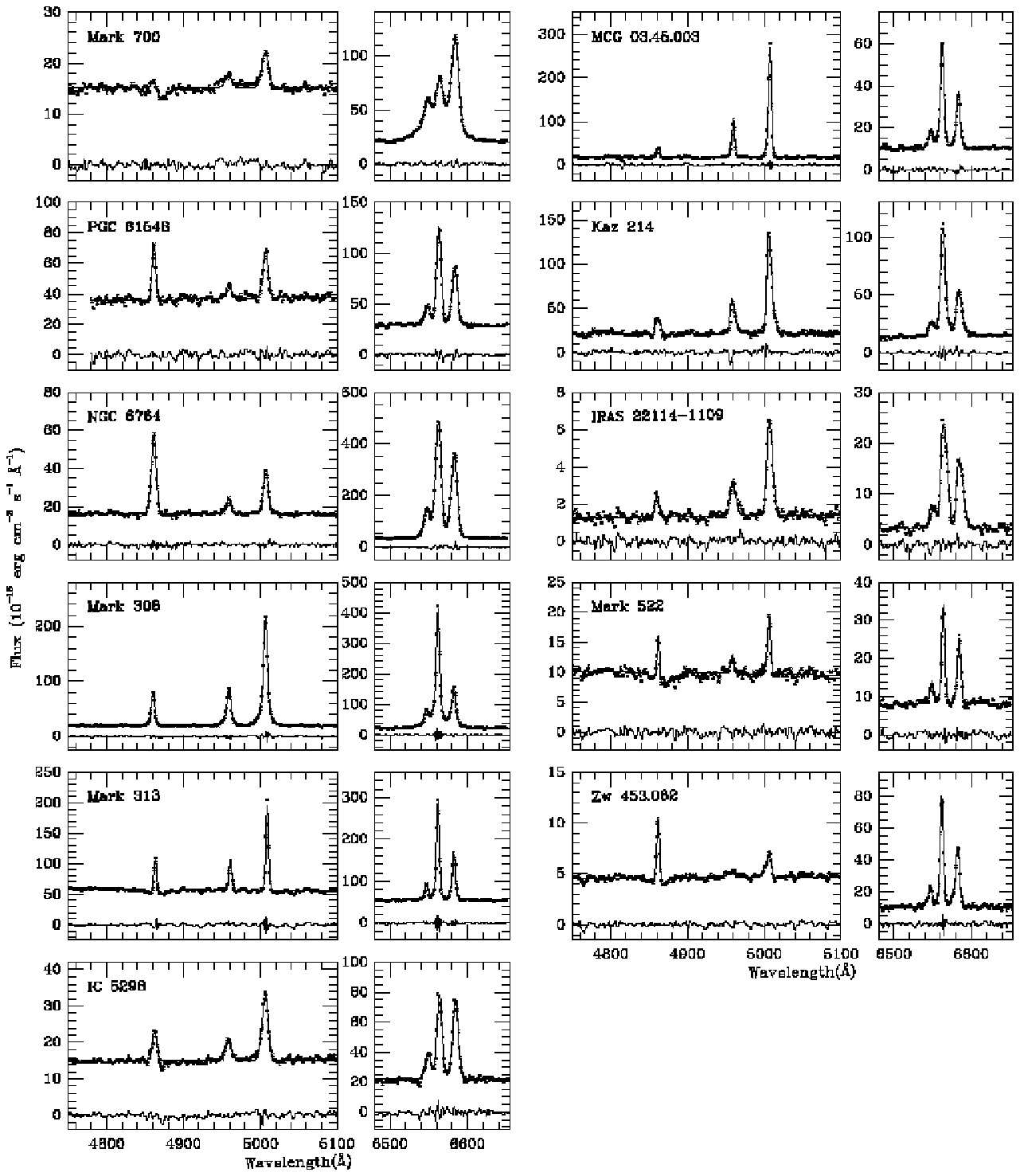}}
\caption{Blue and red spectra for the 53 galaxies studied in this paper (end).}
\end{figure*}

{\bf Mark 1066} is an early-type spiral galaxy (Afanasjev et al. 1981; 
Mazzarella \& Boroson 1993). It was classified as a Seyfert 2 by Afanasjev et 
al. (1980) and as a Seyfert 1.9 by Afanasjev et al. (1981) on the basis of 
weak broad components in the Balmer lines; the existence of these broad 
components has not been confirmed by Goodrich \& Osterbrock (1983) who 
concluded, on the basis of the line ratios (\roiii\ = 4.35, \rnii\ = 0.88, 
\roi\ = 0.08), that 
this object is a Seyfert 2. However, Osterbrock \& Dahari (1983), on the 
basis of the same data, called it a ``marginal'' Seyfert 2, the \roiii\ 
ratio being relatively weak. The spectra 
published by Wilson \& Nath (1990) and Veilleux (1991a) show that the 
emission lines have a broad blue wing extending up to velocities of 
1\,000 \kms\ with respect to the line peaks, the \oiii\ lines being 
significantly broader (403 \kms\ FWHM) than the Balmer lines (280 \kms) 
(Veilleux 1991b,c). De Robertis \& Osterbrock (1986) noted a good correlation 
between the width of the line at half-peak intensity and the critical 
density, suggesting that the narrow line region (NLR) is density stratified; 
however, the density 
stratification mostly affects the high velocity gas producing the wings of 
the line profiles (Veilleux 1991c). Haniff et al. (1988) have published an 
\oiii\ image suggestive of a double structure with a separation of 
0\farcs8, and 
the continuum nucleus in between. There are two emission peaks in the 
core of the low-ionization lines; these peaks are separated by about 0\farcs5, 
the velocity difference between them being $\Delta$V = 125 $\pm$ 20 
\kms\ (Veilleux 1991c). High-resolution (0\farcs1) HST images (Bower et al. 
1995) reveal that the \ha\ and \nii\ emission comes from a 3\arcsec -long 
region centered on the nucleus, while the \oiii-emitting gas is concentrated 
in a bright ``jet-like'' structure extending 1\farcs4 NW of the nucleus. 
Long-slit spectroscopy suggests the existence of two kinematically distinct 
regions: the first, of low-excitation, lies in the plane of the galaxy and 
is normally rotating, while the second, of high-excitation, would be 
inclined with respect to the disk. Bower et al. (1995) suggested that the 
high-ionization cloud is a Seyfert 2 (with \roiii\ $\sim$ 10--15) and the 
low-ionization cloud is a Liner (\roiii\ $\sim$ 2--3); it seems, however, that 
the \oi\ lines are weak and that this region could be a \hii\ region instead. 
Three sets of lines are needed to fit our spectra; one set originates, 
most probably, from a \hii\ region, the two others having line ratios typical 
of Seyfert 2 clouds. The velocity difference between the two Seyfert components 
is $\Delta$V = 146 \kms\ in the blue and 105 \kms\ in the red. The complexity 
of the line emission regions in this object, revealed by the HST observations, 
makes the measured line ratios for each individual component rather 
inaccurate. It seems likely that the density-stratified cloud emitting 
relatively broad lines is compact and coincides with the nucleus. This is, 
therefore, a ``composite-spectrum object''.

{\bf IRAS 03355\,+\,0104} has been identified by de Grijp et al. 
(1987) with a galaxy shown to be a Seyfert 2 by de Grijp et al. (1992) who 
have measured \rnii\ = 0.58, a normal value for such an 
object; however, Vogel et al. (1993) have found much weaker \nii\ lines, 
with \rnii\ = 0.18. Our red spectrum gives \rnii\ = 0.49, in agreement with 
de Grijp et al. (1992), and \roi\ = 0.12, so this object 
is a Seyfert 2 galaxy. In addition, a weak broad \ha\ component 
seems to be present, in which case it would be a Seyfert 1.9 galaxy. 

{\bf IRAS 04210\,+\,0400} has been identified with a compact blue galaxy with a 
faint blue, spiral companion (Moorwood et al. 1986). It is associated with a 
double lobed radio source, 20--30 kpc in size (Beichman et al. 1985; Hill et 
al. 1988) . The galaxy has an apparent spiral structure (Beichman et al.); 
however, these features are dominated by emission lines, and the galaxy 
is probably an elliptical (Hill et al. 1988; Steffen et al. 1996).

We have searched the {\it Hubble Space Telescope} archives and found images 
obtained with the Wide Field Planetary Camera 2, on January 31, 1995 
through medium and broad band filters isolating several emission lines and 
a line-free continuum. We retrieved and analysed these unpublished images, the 
{\it HST} observing log being given in Table \ref{hst_log}. 
The galaxy was imaged on 
the Planetary Camera, a 800 $\times$ 800 pixels CCD with a readout noise 
of $\sim$ 5 e$^{-}$\,pixel$^{-1}$. The pixels size is 15 $\times$ 15 $\mu$m, 
which corresponds to 0\farcs0455 on the sky; the field is 36\farcs4 $\times$ 
36\farcs4 (Trauger et al. 1994; Holzman et al. 1995).

\begin{figure}
\resizebox{8.8cm}{!}{\includegraphics{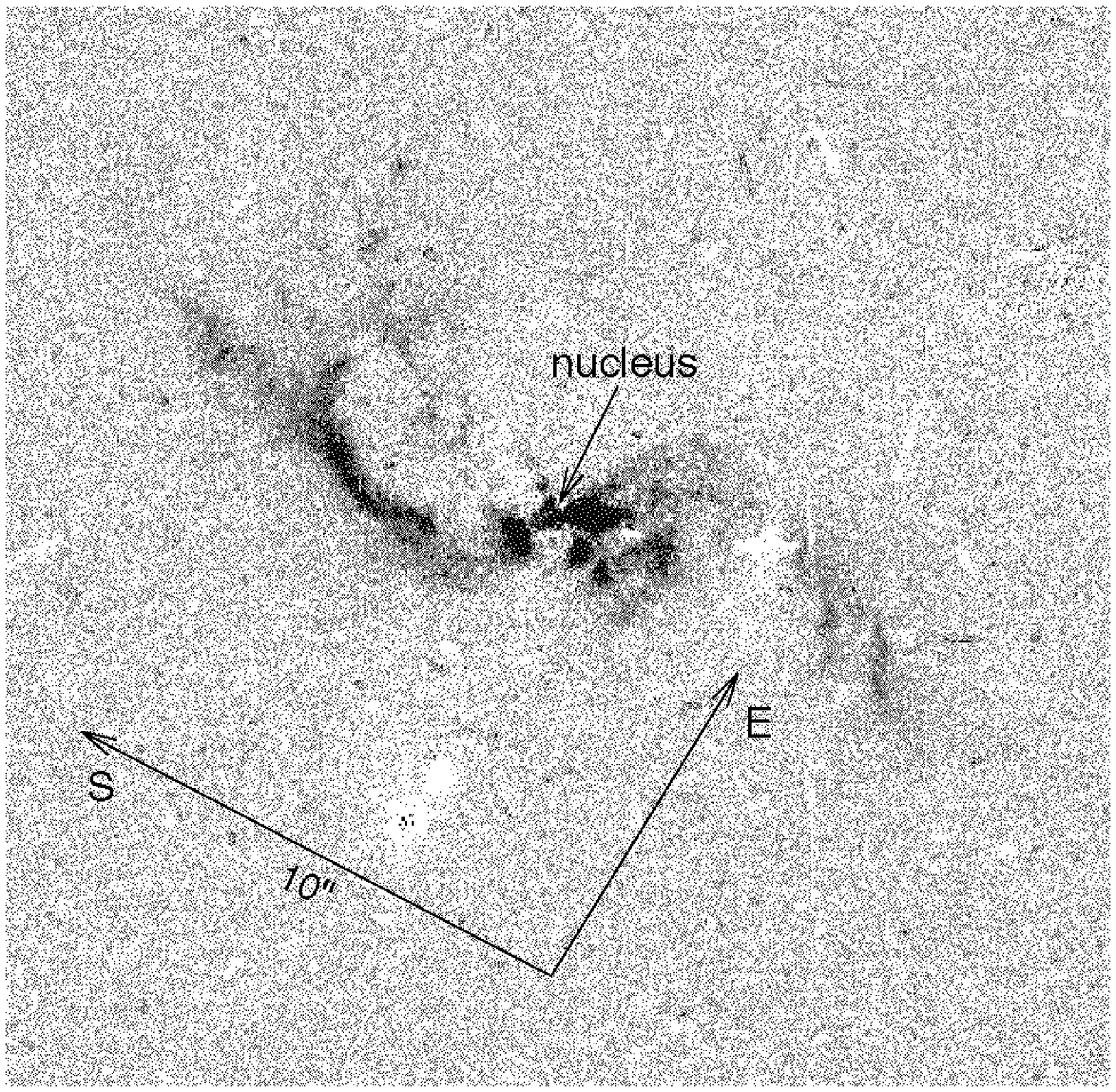}}
\caption{\label{i04210_hst}
\ha+\nii\ {\it HST} image of IRAS 04210$+$0400, after removal of the 
continuum. The nucleus is indicated by an arrow. 10\arcsec, given as a 
reference on the image, correspond 
to 13.5 kpc at the distance of the 
galaxy (assuming $H\rm_{o}$ = 50 \kms\ Mpc$^{-1}$).}
\end{figure}

Both the \ha+\nii\ and the \oiii\ images (after subtraction of the 
continuum) show a 
very complex structure with a bright unresolved nucleus, a relatively bright 
elongated central region extending over $\sim$ 2\farcs4, made of several 
distinct clouds, and a thin spiral feature with a total extent of about 
15\arcsec\ (Fig. \ref{i04210_hst}). 

The Balmer decrement observed 
over a 2\arcsec $\times$ 2\arcsec\ area centered on the nucleus is relatively 
large (\ha/\hb\ = 5.5) (Hill et al. 1988), implying a large extinction 
($A\rm_{V}$ $\sim$ 1.4 mag). We cannot exclude the possibility that the 
extinction varies over the emission nebulosity; therefore, the ratio 
(\ha $+$ \nii)$/\lambda$5007, which is approximately equal 
to \ha$/ \lambda$5007 as \nii$\lambda$6583 $\sim$ 1/3 \ha, cannot 
be taken as an estimate for the excitation 
parameter \roiii. Our entrance aperture (2\farcs1 $\times$ 5\farcs0), with 
the slit oriented in the E-W direction, basically includes the central point 
source and the bright central nebulosity. 

Published nuclear line ratios (Hill et al.) led to the classification of 
this object as a Seyfert 2 (Beichman et al.1985; Holloway et al. 1996) 
although \rnii\ = 0.35, a low value for this class. The core region shows 
asymmetric spatial structure with several separate 
components in velocity and space; there are systematic shifts between peak 
positions for the different lines (Holloway et al. 1996). Our spectra 
basically confirm the line ratios obtained in the nuclear region. This is a 
Seyfert 2 galaxy with abnormally weak $\lbrack \rm NII \rbrack$ emission lines.

\begin{table}[h]
\begin{center}
\caption{\label{hst_log}
{\it HST} observing log of IRAS 04210$+$0400.}
\begin{flushleft}
\begin{tabular}{llcrll}
\hline
ID label  &  Band  & $\lambda\rm_{c}$ & $\Delta\lambda$ & Filter & Time \\ 
          &        &     (\AA)        &      (\AA)      &        & \verb+ +(s)\\
\hline
u2mg0401t & \oiii     &  5479         &       486	& F547\,M & ~300 \\
u2mg0402t & \oiii     &  5479         &       486 	& F547\,M & ~300 \\
u2mg0403t & continuum &  7940         &      1531       & F814\,W & ~600 \\
u2mg0404t & \ha+\nii  &  6814         &       877 	& F675\,W & ~600 \\
\hline
\end{tabular}
\end{flushleft}
\end{center}
\end{table}

{\bf IRAS 04507\,+\,0358} has been identified with an elliptical galaxy 
(de Grijp et al. 1987) shown to be a Seyfert 2 by de Grijp et al. (1992) who, 
however, have measured \rnii\ = 0.28, a very low value 
for an object of such a class. Our red spectrum gives \rnii\ = 0.47 with, 
perhaps, a weak broad Lorentzian \ha\ component. Therefore, this is 
not a ``composite-spectrum object'', but rather a Seyfert 2 galaxy or perhaps a 
Seyfert 1.9, if the broad component is confirmed.

{\bf VII Zw 73} has been classified as a 
Seyfert 2 galaxy by de Grijp et al. (1992) on the basis of its emission line 
ratios (\roiii\ = 3.96, \rnii\ = 0.56); 
however, as in the case of Mark 1066, the \oiii$\lambda$5007 line is rather 
weak for a Seyfert 2. Our blue spectrum shows \hb\ to be clearly narrower 
than the \oiii\ lines, an indication of the probable 
``composite'' nature of this galaxy. To obtain a good fit 
to our blue and red spectra, three components are needed. However, two of 
these components turned out to have similar velocities and widths, making 
the determination of the line ratios rather uncertain. We therefore re-observed 
this galaxy with a higher resolution (33 \Am) in the red on 
October 29, 1997 and in the blue on October 30. Again three components were 
needed to fit the spectra, but this time they were clearly 
identified as corresponding to two Seyfert 2 nebulosities and a 
\hii\ region.

{\bf Mark 622}. The line ratios published by Shuder \& Osterbrock (1981) show 
that it is a Seyfert 2 galaxy, although the \oi$\lambda$6300 relative flux is 
too low for such a class (\roi\ = 0.06). Furthermore, 
these authors found that the \oiii\ lines are much 
broader (FWHM $\sim$ 1\,050 \kms) than the \ha, \nii\ and \oii\ 
li\-nes (FWHM $\sim$ 350 \kms) suggesting the presence of two emission-line 
regions of different ionization. Wilson \& Nath (1990) have shown that, 
in the nucleus, the \nii\ lines are broader than \ha\ (FWHM = 340$\pm$20 
and 240$\pm$20 \kms\ respectively); moreover, the \oiii\ image of this object 
is only slightly resolved, while $\sim$ 60\% of the \ha\ emission comes from 
an extended component (Mulchaey et al. 1996). Our spectra confirm these 
results; in fact, two components are needed in order to obtain a good fit: 
one is representative of a \hii\ region, the other is typical of a 
Seyfert 2 cloud.

{\bf 3C 198.0}. The line ratios in this elliptical radio galaxy 
are those of a \hii\ region, as discussed in Paper I. Our measurements are 
in agreement with the published values: $\lambda$6300$/$\ha\ = 0.05 
and \rnii\ = 0.28. This is therefore a really puzzling 
object.

{\bf KUG 0825\,+\,248}. The published emission-line relative intensities 
(Augarde et al. 1994) are typical of a \hii\ region except for the 
\oi\ lines, which are too strong ($\lambda$6300$/$\ha\ = 0.13). On our 
red spectrum, we measured $\lambda$6300$/$\ha\ = 0.02. Our values 
for \roiii\ and \rnii\ are in agreement 
with the published values. This object is, therefore, a typical \hii\ region.

{\bf IRAS 09111\,-\,1007} has been identified with the 
western component of a galaxy pair (Murphy et al. 1996). The eastern 
component, separated by 40\arcsec, is called IRAS 09111$-$1007\,E, although 
it is probably unrelated to the IR source; it has a ``transition'' spectrum 
with \roiii\ = 3.94, \rnii\  = 0.74 and \roi\ = 0.07 (Duc et al. 1997). Our 
observations suggest that the spectrum of this object is indeed 
``composite'', the \oiii\ lines being broader than \hb. The 
blue spectrum CCD image clearly shows the \hb\ line to be double and spatially 
resolved. In fact, to obtain a good fit to the \ha+\nii\ and \hb+\oiii\ 
lines, three components are needed: one with line ratios typical of a 
Seyfert 2 cloud, and the two others typical of \hii\ regions.

{\bf Zw 238.066}. On the basis of the published line 
intensity ratios, Veilleux et al. (1995) have called this object a Liner; 
however, the \oi\ lines are weak (\roi\ = 0.07). 
Our observations suggest that it has a ``composite'' nature. 
The red spectrum image clearly shows the \ha\ and \nii\ lines to be extended, 
with a low relative intensity of the \nii\ lines. The blue spectrum shows that 
the \oiii\ lines have a broad blue wing not seen in \hb. We have therefore 
fitted both the red and blue spectra with three sets of lines. The blue 
spectrum has a rather poor signal-to-noise ratio which probably explains 
why the parameters of the blue and red fits (especially the line widths) are 
not in good agreement; however, the differences of profile of the 
different lines leave no doubt about the ``composite'' nature of this 
spectrum. Indeed, the fits show that the emission-line spectrum is dominated 
by two \hii\ clouds 
separated by $\sim$ 220 \kms; but there is, in addition, a weak Seyfert 2 
nebulosity with relatively broad lines ($\sim$ 1\,030 \kms\ FWHM). 

{\bf UGC 5101}. This is an ultra-luminous infrared galaxy 
($L\rm_{IR}$ $\geq$ 10$^{12}$ $L\rm_{\sun}$); it is a peculiar galaxy with 
a large ring and a 15 kpc tidal tail extending to the west, which suggests a 
merger, although no companions are known to this galaxy. It has a single 
bright nucleus (Sanders et al. 1988). Optical spectra have been 
published by Sanders et al. who classified it as a Seyfert 1.5 galaxy on the 
basis of a relatively broad \ha\ line, by Veilleux et al. (1995) and Wu et al. 
(1998a,b) who called it 
a Liner, and by Liu \& Kennicutt (1995). However, the published line ratios 
(\roiii\ = 2.9--4.7, \rnii\ = 1.17--1.35, \roi\ = 0.06--0.09) rather indicate a 
``transition'' spectrum. Our red slit spectrum (PA = 270\degr) 
shows spatially extended narrow emission lines with a steep velocity gradient 
across the nucleus in addition to broader, spatially unresolved lines in 
the nucleus itself. We have extracted five columns centered on 
the nucleus and analysed separately the five spectra. In each case, we have 
found a narrow component (FWHM $\sim$ 100--150 \kms) 
with \rnii\ $\sim$ 0.50 and \roi\ $\sim$ 0.05, the velocity decreasing 
from $\sim$ 250 (West) to $-$120 \kms\ (East). 
On three spectra, we detected relatively broad (FWHM $\sim$ 500 \kms) 
lines, with \rnii\ $>$ 1 and \roi\ $<$ 0.40. In addition, on the 
two central spectra, there is a weak, broad (FWHM $\sim$ 1\,200 \kms) 
\ha\ component. It was not possible to perform such a detailed analysis on 
the blue spectrum, which has a much lower signal-to-noise ratio due 
to a large Balmer decrement (\ha/\hb\ = 8.3; Sanders et al. 1988). 
Nevertheless, we can draw some conclusions: the medium width 
\ha\ component flux, coadded on the three central columns, represents 
24\% of the total \ha\ flux on the same three columns 
(excluding the broad \ha\ component); assuming that the Balmer 
decrement is the same for the narrow and medium width components, and that 
the narrow component is a \hii\ region (this component 
having \rnii\ $\sim$ 0.50, must have \roiii\ $<$ 2), we conclude that the 
intermediate width lines set must have \roiii\ $>$ 3.5 and is, therefore, a 
Seyfert 2 cloud. So, UGC 5101 has a ``composite'' spectrum with a rather strong 
starburst component and a Seyfert 1.8 or 1.9 nucleus 
(we are not able to decide between 1.8 or 1.9 as, because of the rather 
poor signal-to-noise ratio around \hb, it is not possible to put a significant 
upper limit to the broad \hb\ component flux). 

{\bf NGC 2989} is a Sc galaxy (Sandage \& Bedke 1994). Published data (Table 
\ref{line_ratios}) indicate a ``transition'' spectrum; Phillips et al. (1983) called it a 
intermediate object, but V\'eron-Cetty \& V\'eron (1984), on the basis of 
the same line ratios, classified it as a \hii\ region. Our measured flux 
ratios (\roiii\ = 1.70, \roi\ = 0.04 and \rnii\ = 0.53) are in good agreement 
with Phillips et al. and show that this is indeed a pure \hii\ 
region with, perhaps a marginally large \rnii\ ratio. 

{\bf CG 49}. This galaxy has been shown to have a Seyfert 2 spectrum 
by Salzer et al (1995); however, they have measured a low relative 
intensity for the \nii\ lines (\rnii\ = 0.30) as the redshifted wavelength 
of \nii$\lambda$6583 is 6873 \AA\ and falls on the atmospheric B band. 
After correction for this absorption, our spectrum gives \rnii\ = 0.80, 
a normal value for a Seyfert 2 galaxy.

{\bf NGC 3185} is a SBa galaxy (Sandage \& Bedke 1994). Its 
emission-line spectrum is power-law photoionized 
according to Stauffer (1982). Ho et al. (1997a) called it a Seyfert 2, 
although the published line ratios indicate a ``transition'' spectrum. 
We have fitted the lines with two sets of Gaussian profiles: 
one system is a \hii\ region; the other corresponds to a Seyfert nebulosity, 
although the \oi\ lines are quite weak (\roi\ = 0.04). 
 
{\bf Arp 107\,A} is the SW component of an interacting galaxy pair (Arp 1966). 
It exhibits Seyfert 2 activity and its spectrum shows very strong \nii\ 
lines (\rnii\ = 3.0) according to Keel et al. (1985). Our spectrum shows a 
more normal value. 
In fact, the lines have a complex profile which can best be fitted by 
two sets of Gaussians having different widths, but similar \rnii\ ratios, 
namely 1.41 and 1.23. 

{\bf NGC 3504}. This Sb galaxy (Sandage \& Bedke 1994) has a ``composite'' 
nucleus showing both non-thermal activity and recent star formation, the 
optical spectrum being dominated by the regions of stellar activity (Keel 
1984). Ho et al. (1993a), who published line ratios for this object, 
suggested that it could be a ``transition'' between a \hii\ region and a Liner, 
but Ho et al. (1997a) called it a \hii\ region. Fitting our red spectrum 
with a single set of Gaussians does not give a satisfactory fit and indicates 
the presence of weak broad wings in the \nii\ lines; two sets of Gaussians 
are needed, revealing the presence of a weak Seyfert-like nebulosity with 
\rnii\ = 1.87. The blue spectrum was also fitted by two sets of Gaussians 
satisfying, respectively, \roiii\ = 0.1 and \roiii\ = 10. The NGC 3504 
spectrum is therefore ``composite'' and dominated by a starburst; a weak Seyfert 
feature is also present. No component showing Liner characteristics was 
detected in this object.

{\bf IRAS 11058\,-\,1131}. In Paper I we concluded, on the basis of a red 
spectrum, that this object, classified as a Seyfert 2 by various authors 
(de Grijp et al. 1992, Osterbrock \& de Robertis 1985), has 
weak \nii\ lines. Re-analyzing the red spectrum, we have found 
the possible presence of a weak broad (FWHM $\sim$ 2\,100 \kms) 
\ha\ component with 24\% of the total \ha\ flux. Our blue 
spectrum confirms that this is indeed an AGN with \roiii\ = 7.6 and 
a relatively strong \heii\ $\lambda$4686 line (\rheii\ = 0.18). 
Our conclusion is that IRAS 11058$-$1131 belongs to the relatively rare 
class of Seyfert 2 galaxies having weak \nii\ lines, discussed in 
Sect. \ref{weak_NII_lines}; 
other galaxies belonging to this class are UM 85 and 3C 184.1 (Paper I).

{\bf NGC 3642} is an Sb galaxy (Sandage \& Bedke 1994). It was classified as a 
Liner by Heckman (1980). Filippenko \& Sargent (1985) noted the presence 
of very narrow emission lines (FWHM $\sim$ 110 \kms) superposed on weak, 
significantly broader components; a weak broad \ha\ component was also 
detected. Koratkar et al. (1995) confirmed the presence of the broad \ha\ 
component and called this object a Liner, although their published line ratios 
rather point to a ``transition'' spectrum. Our spectra confirm the 
presence of a broad \ha\ component (FWHM $\sim$ 2\,160 \kms) and show, in 
addition, that the narrow line spectrum is made of two clouds, one with 
line ratios typical of a \hii\ region and the other of a Seyfert 2 cloud, this 
object being, therefore, a ``composite''.

{\bf Mark 1291}. Spectroscopic observations of this barred 
spiral by Kollatschny et al. (1983) show it to be a ``transition object'' 
between Seyfert 2s and \hii\ regions; however, high excitation lines 
characteristic of Seyfert 2s such as \fexiv\,$\lambda$5303  
and \fex\,$\lambda$6374 are detected. The nuclear emission 
is compact (Gonzalez Delgado et al. 1997). 
%
%
Our optical spectra show a weak broad (FWHM $\sim$ 1\,950 \kms) \ha\ component 
suspected by Kollatschny et al. The narrow lines are well fitted with a 
single Gaussian profile ($\sim$ 160 \kms\ FWHM). The line ratios are 
\roiii\ = 3.84, \roi\ = 0.04 and $\lambda$6583/ \ha\ = 0.73 
(significantly larger than the published value, 0.48); the classification 
of this object is therefore ambiguous: it is a Seyfert 2 in the the \roiii\ 
{\it vs.} \rnii\ diagram and a \hii\ region in the \roiii\ {\it vs.} \roi\ 
diagram. It would be of interest to obtain high-resolution, high 
signal-to-noise spectra of this object to confirm its ``transition'' nature.

{\bf IRAS 11285\,+\,8240\,A} has been classified as a Seyfert 2 gala\-xy 
by Klaas \& Elsasser (1991), with \roiii\ = 8.25 and \rnii\ = 0.46. Our 
red spectrum is well fitted by a single set of components with \rnii\ = 0.45, 
in excellent agreement with the published value, and \roi\ = 0.11. This is 
another example of Seyfert 2 galaxy with marginal\-ly weak \nii\ lines.

{\bf Mark 739} has a double nucleus (Petrosian et al. 1978). 
The eastern nucleus has a Seyfert 1 spectrum (Petrosian et al. 1979; 
Netzer et al. 1987). The western component, Mark 739\,W, has a starburst 
spectrum according to Netzer et al.; however, Rafanel\-li et al. (1993) 
have remarked that \hb\ is unresolved with a resolution of 130 \kms, 
while the \oiii\ lines are significantly broader 
($\sim$ 300 \kms\ FWHM). Our observations show that this spectrum is 
indeed ``composite'' with a Seyfert 2 nucleus and a \hii\ region and, in 
addition, a weak broad \ha\ component.

{\bf SBS 1136\,+\,594} is a Seyfert 1.5 galaxy (Markarian et al. 
1983; Martel \& Osterbrock 1994). The narrow-line spectrum, however, has 
very weak \nii\ lines (\rnii\ = 0.25; Martel \& Osterbrock); 
this is confirmed by our spectra, with even weaker 
\nii\ lines (\rnii\ = 0.10). This object belongs to the class of AGNs 
with very weak \nii\ lines discussed in Sect. \ref{weak_NII_lines}. 
The individual components used to fit the red spectrum of this 
object are plotted in Fig. \ref{spectra}. 

{\bf NGC 3994} is an Sbc galaxy (Sandage \& Bedke 1994) in a triple system, 
interacting both with NGC 3991 and NGC 3995. Based on the observed line 
ratios, Keel et al. (1985) classified it as a 
Liner; the \roiii\ and \rnii\ values rather suggest a \hii\ region. We found 
that its spectrum is ``composite'', the main contribution to the Balmer lines 
coming from a starburst. The relative 
strength of the \oi\ lines is large in the other component (\roi\ = 0.53), 
suggesting that it is a Liner.
 
{\bf NGC 4102} is a Sb galaxy (Sandage \& Bedke 1994). For Ho et. al. 
(1997a) it is a \hii\ region, although its UV spectrum does not resemble that 
of a starburst galaxy (Kinney et al. 1993). The spectrum published by Ho et al. 
(1995) rather indicates a ``transition object''. Our blue spectrum shows 
a \hb~line obviously narrower than the \oiii\ lines, indicating that the 
spectrum is ``composite''. The blue spectrum was fitted with two sets of three 
Gaussians. The broadest \hb\ profile in the fit contains 7\% of 
the total \hb\ flux. The red spectrum having a very high signal-to-noise 
ratio, we needed three sets of three Gaussians to get a good fit; 
we forced one set to have the same width as the broadest 
set in the blue fit. The fitting analysis gives two strong narrow 
components and a weak broad one, containing $\sim$ 5\% of the total 
\ha\ flux and having \rnii\ = 1.57. 
There was no need to use two narrow components to fit 
the \hb\ line, as the spectrum was taken under poor transparency conditions 
and its signal-to-noise ratio is much lower. Our conclusion is that the nucleus 
of NGC 4102 is dominated by a starburst, but that a weak Seyfert 2 component 
is present and detected mainly by the broadening of the \oiii\ lines.

{\bf IRAS 12474\,+\,4345\,S}. For de Grijp et al. (1992), this object is a 
\hii\ region; however, the published \roiii\ line ratio (2.93) is slightly 
high for the corresponding \rnii\ ra\-tio (0.42). Fitting our spectra with 
a single set of lines gives an unsatisfactory result, while the solution 
with two sets of lines is acceptable, with small residuals. One set of 
lines corresponds to a \hii\ region, while the other has \roiii\ = 4.43, 
\roi\ = 0.03 and \rnii\ = 0.40; these values are intermediate between 
those corresponding to \hii\ regions and Seyfert 2 nebulosities. Moreover, 
the \heii\,$\lambda$4686 line is detected with \rheii\ = 0.17, if all the 
\heii\ flux is attributed to the ``transition'' component. We are unable 
to conclude concerning the nature of this second component.

{\bf UGC 8621} is a Seyfert 1.8 galaxy according to Osterbrock \& Martel 
(1993). Our blue spectrum shows the \hb\ line to be much narrower 
($<$ 80 \kms) than the \oiii\ lines ($\sim$ 665 \kms), 
but no evidence of a broad component. To account for the different 
widths observed, we fitted the blue spectrum with two sets of lines; for 
the narrow component, we found \roiii\ = 0.28, while we have 
forced the broader component to have \roiii\ = 10. We fitted the red 
spectrum with two sets of three Gaussians, imposing 
to one of them to have the same width as the broader \oiii\ component; an 
additional Gaussian was added to fit the broad \ha\ wings. 
We find no trace of \oi\ emission 
(\roi\ $\leq$ 0.01) for the narrow component and \roi\ $<$ 0.3 for the 
Seyfert cloud. This is a ``composite object'', with a relatively strong \hii\ 
region and a weak Seyfert 1.9 nebulosity.

{\bf Mark 266} is a merging system with 
two nuclei separated by 10\arcsec\ (Hutchings \& Neff 1988; Wang et al. 1997). 
It is a Luminous Infrared Galaxy (LIG), i.e., 
10$^{11.2}$ $< L\rm_{IR} <$ 10$^{12}$ $L_{\sun}$ (Goldader et al. 1997). 
Line intensity ratios in the two nuclei have been measured by Osterbrock \& 
Dahari (1983), Veilleux \& Osterbrock (1987), Mazzarella \& Boroson (1993), 
Osterbrock \& Martel (1993) and Kim et al. (1995); all these measurements 
are in good agreement, if we make the assumption that Kim et al. have 
inverted the NE and SW components. On the basis of these line ratios, 
Mark 266\,SW has been classified as a Seyfert 2 by Mazzarella \& Boroson, 
Kim et al. and Wu et al. (1998b), and as a ``marginal'' Seyfert 2 by 
Osterbrock \& Dahari, while Mark 266\,NE has been called a Liner by 
Mazzarella \& Boroson, Kim et al. and Wu et al., and a 
``narrow emission-line galaxy'' (NELG) by Osterbrock \& Dahari and 
Veilleux \& Osterbrock. NELGs, for 
these authors, are emission-line galaxies that may be either Liners or 
\hii\ regions. For Mark 266\,NE, we have only a blue spectrum which, by simple 
visual inspection, shows quite different \hb\ and \oiii\ line profiles. Our 
line profile analysis reveals two clouds, one with narrow lines 
(FWHM $\sim$ 300 \kms) and 
\roiii\ = 0.96, the other with broader lines (FWHM $\sim$ 1\,000 \kms) and 
\roiii\ = 2.2 and containing 37\% of the total \hb\ flux. The published 
value of the intensity ratio \roi\ is 0.15. If the narrow component is 
associated with a \hii\ region, it contributes in a small amount to the 
observed \oi$\lambda$6300 flux and therefore the \roi\ ratio for the broader 
component is likely to be significantly larger than 0.12, which means that this 
component could be a Liner. We have fitted the blue spectrum of Mark 266\,SW 
with three sets of Gaussians: one of them corresponds, most probably, to a 
\hii\ region with narrow lines (295 \kms\ FWHM) and \roiii\ = 0.36; 
the two other sets have widths of 200 and 600 \kms\ and \roiii\ = 3.7 
and 13.3, respectively. We also fitted the red spectrum with three sets 
of Gaussians, 
forcing, in addition, one of the sets to have a width of 600 \kms. The result 
is a set of narrow lines with \rnii\ = 0.58 corresponding to the 
narrow blue lines, confirming that this system is indeed coming from a 
\hii\ region. The set having the broadest lines has intensity ratios typical 
of a Seyfert cloud. The third set, with \roiii\ = 3.65 and \rnii\ = 0.58, has 
still an intermediate spectrum.

{\bf Mark 1361} was called a Seyfert 2 galaxy by Kim et al. (1995). Our 
analysis of a red spectrum (Paper I) lead to the conclusion that it is a 
``composite object''. Our blue spectrum confirms this result. Three sets of 
three components were needed to get a good fit. 
In one set we had to impose 
\roiii\ = 10, in another \roiii = 0.1. The best fit resulted in a narrow set 
of lines with very weak \oiii\ lines and two sets of lines with 
strong \oiii\ contribution. We then re-analyzed the red spectrum using three 
sets of three lines; we had to add a weak broad \ha\ component 
(FWHM $\sim$ 2\,400 \kms, with $\sim$ 10\% of the total \ha\ flux) in 
order to obtain a good fit. For the 
narrowest set of three lines, we found \rnii\ = 0.54, for the other two, 
0.66 and 1.04 respectively. The conclusion is that Mark 1361 has a 
``composite'' spectrum with a \hii\ cloud contributing half of the 
\hb\ line and a Seyfert 2 
nebulosity with complex line profiles (two Gaussians were needed for the fit). 
If the presence of the broad \ha\ component is confirmed, this object could be 
a Seyfert 1.9 galaxy.

{\bf Mark 461} is a Seyfert 2 galaxy according to Huchra \& Burg (1992); 
however, Cruz-Gonzalez et al. (1994) have measured \roiii\ = 1.13. The emission 
is concentrated in the nuclear region (Gonzalez Delgado et al. 1997). 
No other line ratios have been published for this object. The \hb\ and 
\oiii\ lines obviously do not have the same profile. To get 
a good fit, two sets of lines were necessary. The object is ``composite'', one 
component being a \hii\ region and the other probably a Seyfert 2 nucleus.

{\bf 4C 26.42}. This object has been identified with a cD galaxy, 
MCG 05.33.005 (Carswell \& Wills 1967; Olsen 1970; Merkelijn 1972), the 
brightest member of Abell 1975 (Parma et al. 1986; Pilkington 1964); it is 
a FR~I, Z-shaped radiosource (van Breugel et al. 1984; Ge \& Owen 1993). 
Emission lines have been detected in the nuclear region, with \roiii\ = 0.4, 
\rnii\ = 0.8 and \roi\ = 0.2 (Anton 1993). These 
values are similar to the ones usually observed in Liners, but for 
the low \oiii$\lambda$5007 line intensity. Examination of 
the red spectrum shows that the lines are obviously double. Fitting the lines 
with two sets of components revealed two clouds with a velocity difference of 
$\sim$ 330 \kms. Their line ratios are very similar and typical of Liners 
with exceptionally weak \oiii\ lines.
 
{\bf I Zw 81}. Koski (1978) observed narrow (FWHM = 225 $\pm$ 200 \kms) 
emission lines in this galaxy, with $\lambda$5007/\hb\ = 3.14, 
$\lambda$6583/\ha\ = 0.67 and \roi\ = 0.07; the 
narrowness of the emission lines and 
the relative weakness of the \oiii\ lines led him to 
conclude that this is not a Seyfert galaxy, but rather a ``transition'' 
case between a \hii\ galaxy and a Seyfert 2. Shuder \& Osterbrock (1981) 
and Veilleux \& Osterbrock (1987) called it a narrow-line Seyfert. 
Our red spectrum shows the lines to be double with a separation of 220 
\kms. Fitting both the red and blue spectra with two sets of three Gaussians, 
we found \roiii\ = 2.05 (1.41) and \rnii\ = 0.78 (0.56) for the high (low) 
velocity clouds. In both cases, the \oi\ lines are undetected with 
\roi\ $<$ 0.04. The two clouds are most probably \hii\ regions.
  
{\bf Mark 477}. This object was discussed in Paper I, where we concluded from 
the published data and the analy\-sis of a red spectrum that its nature 
was unclear. Heckman et al. (1997) have argued that the observed UV through 
near-IR continuum in the nucleus of Mark 477 is dominated by light from a 
starburst. The narrow Balmer emission lines would then be excited by ionizing 
radiation from both the hidden Seyfert 1 nucleus and from the hot stars in 
the starburst. We have re-analyzed our old red spectrum together with our new 
blue spectrum. Three sets of components were needed to fit these very high 
signal-to-noise data. The three line-systems are typical of Seyfert 2s, 
although two have relatively weak \nii\ intensities, with 
$\lambda$6583/ \ha\ = 0.31 and 0.38, respectively. We found 
no evidence for the presence of broad Balmer components. 

{\bf Mark 848\,S} is a LIG (Goldader et al. 1997) belonging to a pair of 
interacting galaxies (Armus et al. 1990). The northern galaxy is a \hii\ 
region (Wu et al. 1998a,b); the southern component has been called a 
Liner (Mazzarella \& Boroson 1993; Veilleux et al. 1995), though its 
line ratios (Kim et al. 1995) are ambiguous, the \oi$\lambda$6300 line being 
rather weak for a Liner. The line profile analy\-sis of our spectra shows it 
to be a ``composite object'' with two distinct emission-line clouds: a narrow 
one (FWHM $\sim$ 140 \kms, with \roiii\ = 0.83, \rnii\ = 0.43 and 
\roi\ = 0.03), identified with a \hii\ region, and a somewhat broader one 
(FWHM $\sim$ 580 \kms), with line ratios ty\-pical of a Seyfert 2 
(\roiii\ = 4.20, $\lambda$6583/ \ha\ = 0.71 and \roi\ = 0.14).

{\bf IRAS 15184\,+\,0834}. De Grijp et al. (1992) called this object a 
Seyfert 2; however they found relatively weak \nii\ lines (\rnii\ = 0.42). 
Our spectra show that the \nii\ lines are significantly stronger than the 
published values. A reasonable fit is obtained with two sets of components: 
one is a \hii\ region; the other could be a Seyfert 2, although the \oi\ 
lines are relatively weak.

{\bf NGC 5953} is a peculiar S0 (Rampazzo et al. 1995) or Sa 
(Delgado \& Perez 1996) galaxy interacting with NGC 5954 (Arp 1966). It 
has a Seyfert 2 nucleus (Rafanelli et al. 1990; Delgado \& Perez) 
surrounded by a ring of star formation with a radius of $\sim$ 4\arcsec\  
(Delgado \& Perez). Rafanelli et al. and Delgado \& Perez 
studied this object using a slit width of 2\farcs0 and 1\farcs5 respectively. 
The seeing was $\sim$ 1\arcsec\ during Delgado \& Perez' observations, while 
it was not specified by Rafanelli et al. who, however, easily separated 
the galaxy nucleus from a star located 3\arcsec\ away. 
We may reasonably assume that, in both cases, the nuclear spectrum 
corresponds to a 2\farcs0 $\times$ 2\farcs0, or smaller, aperture. 
The line ratios given by these authors (see Table \ref{line_ratios}) are 
typical of a 
Seyfert 2 region, although \oi$\lambda$6300 may be somewhat weak for 
this type of objects; but as stressed before, we cannot exclude some 
contamination by the circumstellar emission region. 
Both Keel et al. (1985) and Kim et al. (1995) have observed the NGC 5953 
nuclear region with a relatively large aperture: Keel et al. used a 
$\phi$ = 4\farcs7 circular aperture, while Kim et al. used a long 
2\arcsec\ $\times$ 14\arcsec\ slit. It is clear 
that these two large aperture spectra must contain a significant 
contribution from the circumstellar emission region and, indeed, the 
published line intensity ratios are those of ``transition'' spectra. 
We used a 2\farcs1 slit and the seeing was 2\farcs6; we extracted 
7 columns, i.e., 7\farcs6, so some 
contamination from the circumstellar emission region was expected. 
Effectively, by doing the line profile fitting analy\-sis, we were able to 
identify two components of different line widths and strengths, one of 
which can be associated with a \hii\ region 
(\roiii\ = 0.55, \rnii\ = 0.60, \roi\ = 0.03 and FWHM $\sim$ 200 \kms); the 
other component, broader (FWHM $\sim$ 400 \kms), reveals the 
presence of a Seyfert 2 nebulosity, the measured line intensities being: 
\rnii\ = 1.96 and \roi\ = 0.18, with \roiii\ fixed to 10. 
Lines of \feiii$\lambda$4659 and \fevii$\lambda$5158 are clearly visible in 
the blue spectrum. A very weak broad \ha\ line is possibly detected, 
which would qualify NGC 5953 as a Seyfert 1.9 galaxy.

{\bf Kaz 49} has been classified as a Seyfert 1 by Yegiazarian \& Khachikian 
(1988), as a Seyfert 1.9 by Moran et al. (1994), who have detected a weak broad 
\ha\ component (FWHM = 1\,150 \kms), and as a \hii\ region by Boller et al. 
(1994). The latter classification was based on measured line ratios 
(\roiii\ = 2.58, \rnii\ = 0.56, \roi\ = 0.025) that rather point to a 
``transition'' spectrum. The line profile analysis of our spectra shows a 
strong \hii\ region (\roiii\ = 2.21, \roi\ = 0.05, \rnii\ = 0.55) and a 
weak Seyfert 2 component for which we have fixed \roiii\ = 10. There is no 
evidence for the presence of a broad \ha\ component; however, the blended 
weak \ha\ and \nii\ components, each having a FWHM $\sim$ 880 \kms\ may be 
easily mistaken for a broad \ha\ line.

{\bf IRAS 16129\,-\,0753} has been classified as a possible Liner by de 
Grijp et al. (1992) on the basis of the measured line intensity ratios 
(\roiii\ = 2.03, \rnii\ = 0.64), although \oi$\lambda$6300 was very weak. 
The line fitting analysis of our blue spectrum shows this object to be 
``composite''. The red spectrum, which has a relatively low signal-to-noise 
ratio, is well fitted by a single set of lines corresponding to the \hii\ 
region; the Seyfert component is undetected.

{\bf IRAS 16382\,-\,0613} has been called a Seyfert 2 by Aguero et al. 
(1995) and a possible Seyfert 2 by de Grijp et al. (1992); however, the 
\oi$\lambda$6300 line is marginally weak for a Seyfert 2, with \roi\ = 
0.09 (Aguero et al.). The line profiles on the blue spectrum are obviously 
complex. Fitting these lines with two sets of Gaussians reveals a narrow 
component (FWHM $\sim$ 350 \kms) with \roiii\ = 3.94, and a broader 
component (FWHM $\sim$ 1\,160 \kms) with \roiii\ = 4.06. The red spectrum 
fit gives a solution compatible with the blue solution plus a broad Balmer 
line (FWHM $\sim$ 5\,000 \kms). The two components have strong \nii\ lines, 
but the \oi\ lines are weak. For the broadest set of lines, we find \roi\ 
$<$ 0.12, compatible with a Seyfert 2 nebulosity; however, the narrow 
component has \roi\ $<$ 0.03 and seems therefore to have a genuine 
``transition'' spectrum. 

{\bf Mark 700} was called a Seyfert 1 galaxy by Denisyuk et al. (1976), who 
found a broad \ha\ component. For Koski (1978), it is a weak-lined Seyfert 
galaxy with Balmer absorption lines, very similar to ``normal'' 
emission-line galaxies. Ferland \& Netzer (1983) included it in a 
Liner list, on the basis of the intensity ratios published by Koski. 
Our observations show that this object is, indeed, a Liner; the 
broad \ha\ component seen by Denisyuk et al. is confirmed. 

{\bf MCG 03.45.003}. The \nii\ lines measured by de Grijp et al. (1992) are 
rather weak for a Seyfert 2 galaxy (\rnii\ = 0.42) and, on the basis of a red 
spectrum, we concluded in Paper I that this object could have a ``composite'' 
spectrum. Our analysis of both the blue and red spectra show 
that two kinematically distinct clouds are present in this object, 
both of them having Seyfert 2 characteristics.

{\bf PGC 61548}. The red spectrum is ``composite'' and confirms the 
result presented in Paper I. The line profile analysis reveals the presence 
of both a \hii\ region ($\lambda$5007/\hb\ = 0.41,  
$\lambda$6583/\ha\ = 0.50, $\lambda$6300/\ha\ = 0.04, FWHM $\sim$ 250 \kms) 
and a Seyfert 2 nebulosity \roiii\ fixed to 10.0, 
\rnii\ $\sim$ 3.9, \roi\ $\sim$ 0.5 and FWHM $\sim$ 570 \kms). 

{\bf Kaz 214} is a Seyfert 2 galaxy for de Grijp et al. (1992), with 
\roiii\ = 5.23 and \rnii\ = 0.39; however, the \nii\ lines are weak for 
a Seyfert 2. On our red exposure, the slit position angle was PA = 139\degr. 
By simple visual inspection of the CCD image, we see that the lines are double: 
in one of the line-systems the lines are spatially extended and narrow, 
with relatively weak \nii; in the other, the lines are spatially unresolved, 
but relatively broad, and \nii\ is stronger. The spectrum is obviously 
``composite'' with a \hii\ region and a Seyfert component. However, when analysing 
the spectrum obtained by extracting three columns centered on the nucleus, 
we were unable to get a satisfactory fit confirming the visual impression. 
We then extracted individually seven columns (numbered 1 to 7, from SE to NW) 
containing obvious emission lines; the continuum was brightest in columns 
4 and 5. Columns 1, 2 and 7 were fitted with a single set of lines, while for 
columns 3 to 6, two sets of lines were necessary. We have made the assumption 
that the Seyfert component is really spatially unresolved and, consequently, 
forced the redshift, width and the \rnii\ ratio of this component to be the 
same on all columns (that is, 120 \kms, 525 \kms\ FWHM and 0.60, 
respectively), the 
only free parameter being the \ha\ intensity. In addition to this Seyfert 
component, we have found, on all columns, a narrow component with relatively 
weak \nii\ lines; the velocity of this narrow component increases from 
$-$25 \kms\ to 140 \kms\ from column 1 to 7. The blue spectrum was taken with 
the slit oriented N-S. As the seeing was rather poor, seven columns were 
added together when extracting the spectrum. The best fit was obtained with 
three sets of lines: for one of them, we forced \roiii\ = 10 (this turns to be 
the broadest component); the two other sets have narrow lines, with moderate 
\roiii\ ratios. We therefore conclude that Kaz 214 has a ``composite'' spectrum. 
But this example shows that it may not be possible to show that 
a ``transition'' spectrum is a ``composite'' spectrum when 
the spatial resolution is insufficient, 
this being due to the large velocity dispersion gradient sometimes present 
in the nuclear region which broadens the lines.

{\bf NGC 6764} has been called a Seyfert 2 galaxy by Rubin et al. (1975), 
in spite of \hb\ being stronger than \oiii$\lambda$5007; this classification 
was based on the width of the \ha\ and \nii\ lines ($\sim$ 750 \kms) 
but Wilson \& Nath (1990) found these lines to be much narrower 
($\sim$ 380 \kms\ FWHM). Koski (1978) noticed the presence of weak 
\hi\ absorption lines and, from the line intensity ratios, concluded 
that it was very much like ``normal'' galaxies, while for Shuder \& 
Osterbrock (1981), it is not a Seyfert 2. 
Using the line ratios published by Koski (1978), Ferland \& Netzer (1983) 
classified it as a Liner. Osterbrock \& Cohen (1982) have detected in the 
spectra of this object the $\lambda$4650 Wolf-Rayet emission feature. For 
Ashby et al. (1992), it is a starburst galaxy. Line profile fitting of our 
spectra revealed the ``composite'' nature of this object. 
Two systems were identified: a narrow one (FWHM = 325 \kms) 
with line ratios compatible with those usually observed in \hii\ regions 
(\roiii\ = 0.62, \rnii\ = 0.65 and \roi\ = 0.04) and a broader system 
(FWHM = 480 \kms) with line ratios similar to those of Liners 
(\roiii\ = 0.44, \rnii\ = 0.96 and \roi\ = 0.14). This is, therefore, 
a ``composite object''.

{\bf IRAS 22114\,-\,1109} was classified a Seyfert 2 by Veilleux et al. 
(1995); however, the \oiii\ lines are relatively weak 
for this type of objects (\roiii\ = 4.22; Kim et al. 1995). 
A line profile analysis was performed on the red and blue 
spectra. The measured line intensities and widths are compatible with the 
simultaneous presence on the slit of both a \hii\ region 
(\roiii\ = 1.33, \rnii\ = 0.70, \roi\ $<$ 0.07 and FWHM = 185 \kms) and a 
Seyfert 2 nebulosity (\roiii\ fixed to 10.0, \rnii\ = 0.60, 
\roi\ = 0.12 and FWHM = 415 \kms), so this is another example of a 
``composite-spectrum object''. 

{\bf Mark 308} was called a Seyfert 2 galaxy by Popov \& Khach\-ikian 
(1980) and Zamorano et al. (1994); V\'eron-Cetty \& V\'eron 
(1986a) classified it as a \hii\ region, although the published 
line ratios (Table \ref{line_ratios}) are unlikely 
for either classes. The analysis of our blue spectrum (Paper I) showed this 
object to be ``composite'' with one narrow component with weak \oiii\ lines and 
two broader components with strong \oiii\ lines; the analysis of our red 
spectrum confirms this result, the narrow system (FWHM $\sim$ 155 \kms) being 
typical of a \hii\ region (with \roiii\ fixed to 0.1, 
\rnii\ = 0.30 and \roi\ = 0.06) 
and the two broader line-sets (FWHMs $\sim$ 325 and 1\,045 \kms, respectively) 
of Seyfert-like clouds. Moreover, we have detected a weak broad 
(1\,725 \kms\ FWHM) 
\ha\ component containing $\sim$ 7\%  of the total \ha\ flux. The companion 
galaxy, KUG 2239$+$200\,A, at $z$ = 0.024 (Keel \& van Soest 1992) and located 
53\arcsec\ NE of Mark 308, has a \hii-like emission-line spectrum.

{\bf Mark 522} is a Seyfert 2 galaxy according to Veilleux \& Osterbrock 
(1987); however, the \oiii\ and \oi\ lines are relatively weak (\roiii\ 
= 3.23, \roi\ =0.07). Our observations show this object to be ``composite'', 
with two different line systems: one, ``narrow'' (FWHM $\sim$ 100 \kms), 
typical of a \hii\ region (\roiii\ = 0.63, \rnii\ = 0.53 and \roi\ $<$ 0.04), 
the other, somewhat broader (FWHM $\sim$ 240 \kms),  associated with a 
Seyfert 2 nebulosity ($\lambda$5007/ \hb\ = 7.87, \rnii\ = 1.50 and 
\roi\ $<$ 0.2).

{\bf Mark 313} is a Seyfert 2 galaxy according to Osterbrock \& Pogge (1987) 
(with \roiii\ = 3.52, \rnii\ = 0.52 and \roi\ = 0.10), and Moran et al. 
(1996); the \oiii\ lines are relatively weak for a Seyfert 2. 
Images in \oiii\ and \ha+\nii\ show a very complex structure, with high 
excitation gas restricted to a symmetric, linear feature (Mulchaey et al. 
1996). From a two-component Gaussian fitting of high-dispersion spectra 
of the nucleus of this object, Maehara \& Noguchi (1988) concluded that 
it is a ``composite object'' with a \hii\ region and a Liner nebulosity. 
Line profile analysis of our spectra reveals the contribution of two 
different line-emitting regions: one, with \roiii\ = 2.29, \rnii\ = 0.44, 
$\lambda$6300/ \ha\ = 0.10 and narrow width (135 \kms\ FWHM), is typical 
of a \hii\ region; the other, much weaker, is not detected in the blue 
and its line ratios are \rnii\ = 0.71 and \roi\ = 0.28; it could be either 
a Seyfert 2 or a Liner, depending on the \roiii\ ratio. 

{\bf Zw 453.062} is a LIG (Goldader et al. 1997); it was called a Liner by 
Veilleux et al. (1995) on the basis of the measured emission line ratios, 
while Wu et al. (1998a,b) found its properties to be intermediate between 
\hii\ regions and Liners, although the \oi\ lines are very weak. Our spectra 
suggest that it is a ``composite object'', one component being a Seyfert 2 
nebulosity and the other a \hii\ region.

{\bf IC 5298} is a LIG (Goldader et al. 1997). Wu et al. (1998a,b) found 
that its properties are intermediate between \hii\ regions and Liners; 
it was classified as a Seyfert 2 by Veilleux et al. (1995) although the 
\oi\ lines are rather weak (\roi\ = 0.05). Our observations suggest 
that the spectrum is ``composite'', being dominated by a \hii\ region. 


\section{Results}

\begin{figure}
\resizebox{8.4cm}{!}{\includegraphics{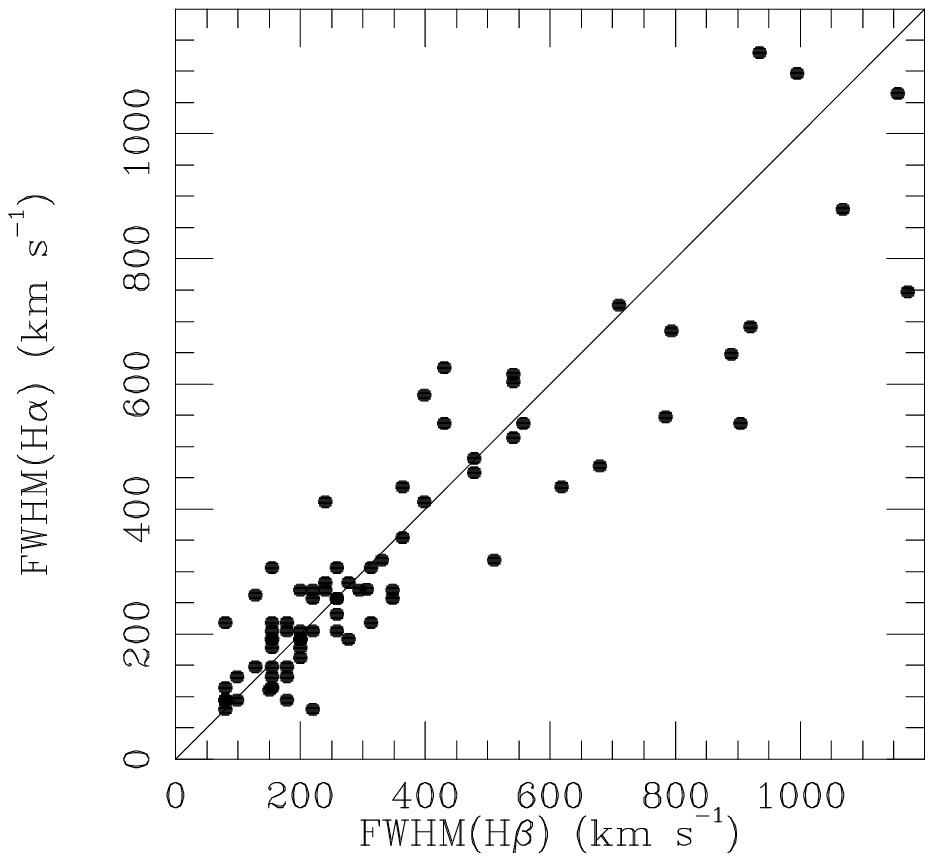}}
\caption{\label{fwhms}
FWHM of all the individual line-components measured on 
the red spectra {\it vs.} the FWHM of the individual components measured on 
the blue spectra.}
\end{figure}

In Fig. \ref{fwhms} we have plotted the FWHMs (corrected for the instrumental 
broadening) of each individual com\-ponent, i.e. of each 
set of lines used to fit the blue and red spectra, as listed in Table 
\ref{fits} (cols. 5 and 10, respectively). The good correlation found 
between the blue and red FWHMs gives confidence in the fitting analysis.

\begin{figure*}
\resizebox{12cm}{!}{\includegraphics{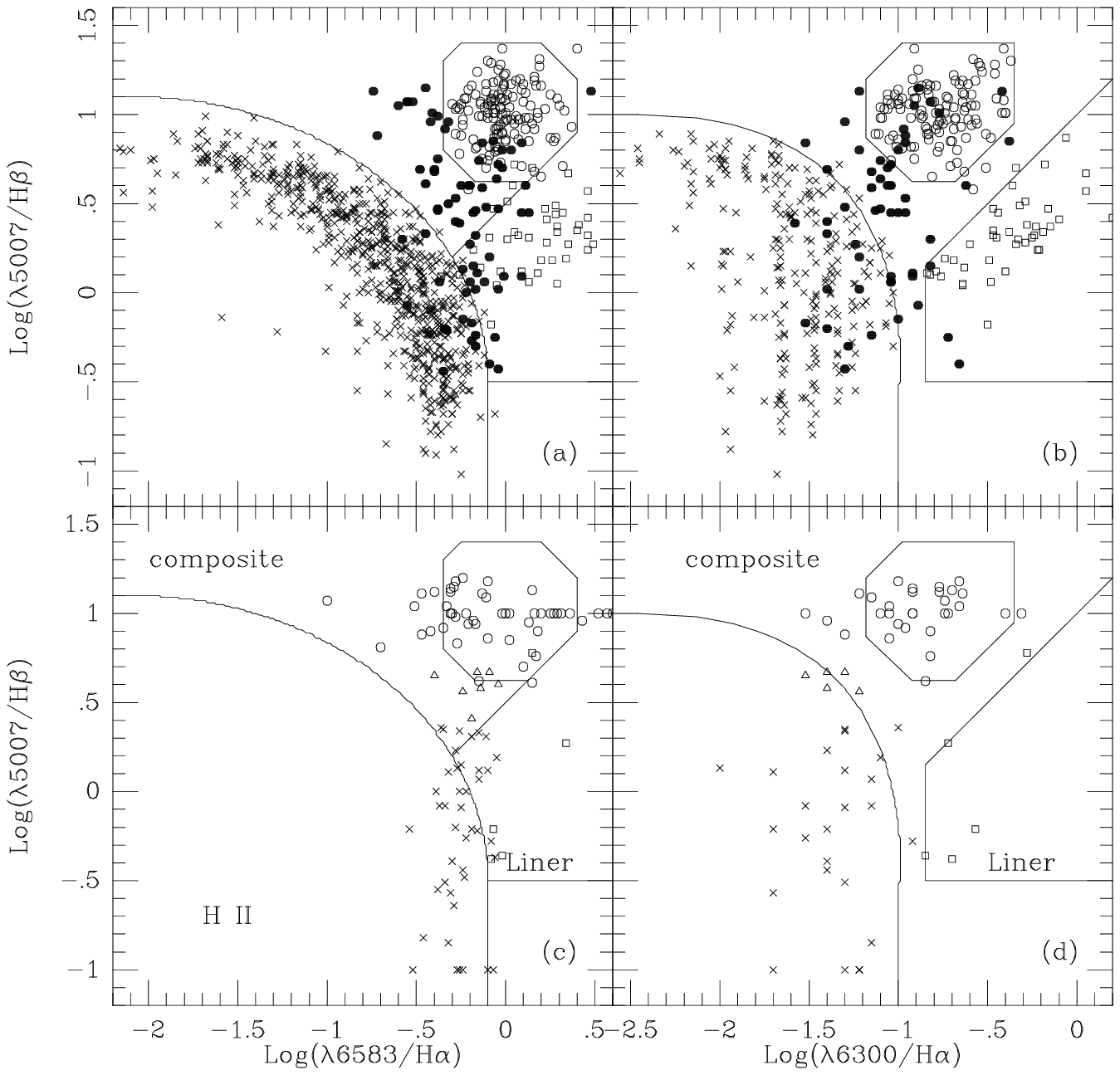}}
\hfill
\parbox[b]{55mm}{
\caption{\label{dd}
Diagnostic diagrams showing the log(\roiii) {\it vs.} log(\rnii) -- 
boxes (a) and (c) -- and log(\roiii) {\it vs.} log(\roi) -- boxes 
(b) and (d). \hii\ regions are represented by crosses, 
Seyfert 2 galaxies by open circles, 
and Liners by open squares. In (a) and (b), filled circles represent 
``transition objects'',  i.e. objects which, in at least one of the 
diagrams, fall outside the arbitrarily delimited regions assigned to 
\hii\ regions, Seyfert 2s and Liners. In (c) and (d) we plotted 
the individual components. The symbols are the same as in the upper panels; 
open triangles represent objects which could not be classified (``?'' in 
Table \ref{fits}).}
}
\end{figure*}

Figure \ref{dd} shows the log(\roiii) {\it vs.} 
log(\rnii) and log(\roiii) {\it vs.} 
log(\roi) diagrams traditionally used to classify nuclear emission-line 
regions into \hii\ regions, Liners or Sey\-fert 2s. We have delimited in 
the two diagrams three regions, each corresponding to one of these classes. 
In Figs. \ref{dd}a and \ref{dd}b we have plotted all 
objects for which line ratios are 
available in the literature and which are unambiguously 
classified as \hii\ regions (crosses), Seyfert 2s (open circles) or 
Liners (open squares); we have also plotted the 61 observed objects 
suspected of having a ``transition'' spectrum (filled circles): 
they fall, at least in one of the diagrams, in a 
``zone of avoidance'', i.e. outside the regions arbitrary assigned to the 
classical emission-line regions. In figures \ref{dd}c and \ref{dd}d, which are 
the same as \ref{dd}a and \ref{dd}b respectively, we have plotted 
the individual components used to fit the spectra, as given in Table \ref{fits}.

It is apparent that most of the 
``transition objects'' belong to one of the three following categories: 
\begin{enumerate}
\item{A few objects fall into the ``zones of avoidance'' only because they 
have inaccurate published line ratios, appearing to be ``normal'' when more 
accurate measurements are obtained; this is the case, for instance, for 
IRAS 04507$+$0358, KUG 0825$+$248, NGC 2989, CG 49 and Arp 107A.} 
\item{A few objects have Seyfert 2 spectra with abnormal\-ly weak \nii\ lines. 
They constitute a rare but interesting class of objects which is further 
discussed below.} 
\item{Most ``transition'' spectra turn out to be ``composite'', 
due to the simultaneous presence on the slit of a \hii\ region and a Seyfert 2 
nebulosity. We have observed 70\% of all the objects in an unbiased sample of 
galaxies displaying a ``transition'' nuclear spectrum. Modeling of the data 
revealed that most of them have in fact a ``composite'' spectrum, 
suggesting that genuine ``transition objects'' do not exist at all. However, 
in a few cases such as NGC 3185, Mark 1291, 
IRAS 12474$+$4345\,S, Mark 266\,SW or IRAS 15184$+$0834, we 
cannot prove that the spectra are ``composite''; the classification 
is ambiguous. Further studies are needed to find out the true nature of these 
``transition objects''.} 
\end{enumerate}

\begin{figure}
\resizebox{8.4cm}{!}{\includegraphics{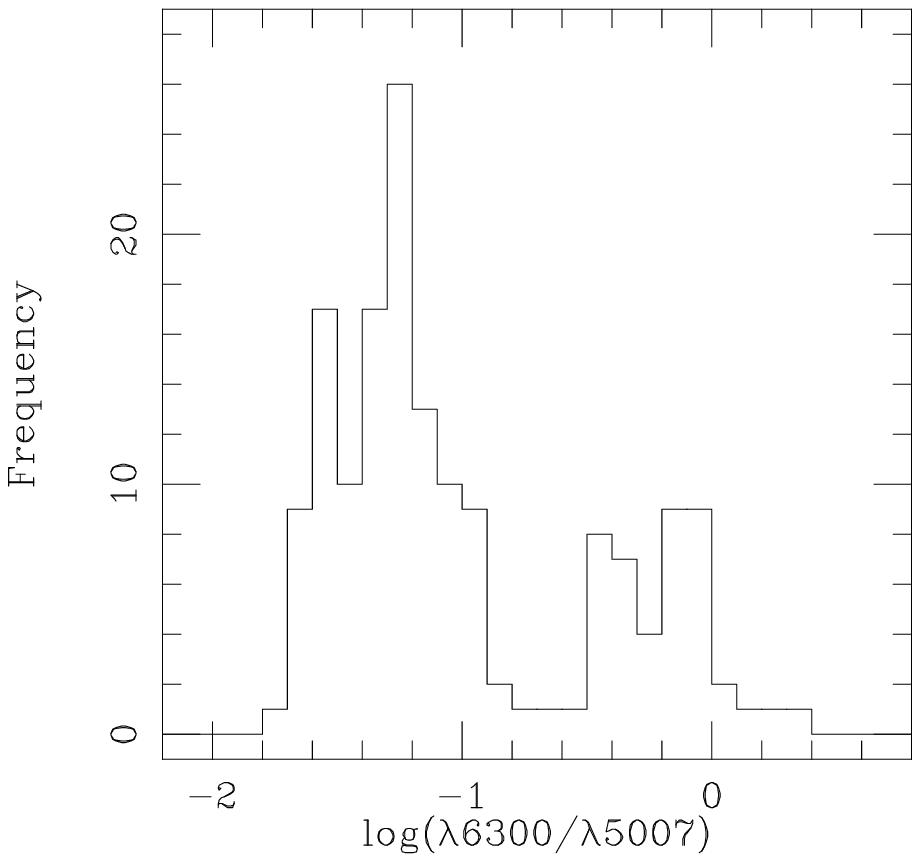}}
\caption{\label{histo_oi_oiii}
Histogram of log(\rliner) for the 159 Seyfert 2 and Liners plotted 
in Fig. \ref{dd}.}
\end{figure}

Fig. \ref{histo_oi_oiii} is the histogram of the parameter log
($\lambda$6300/$\lambda$5007) for 159 Seyfert 2s and Liners after correction 
of the line fluxes for reddening, 
assuming that the intrinsic Balmer decrement is \ha /\hb = 3.1 (Osterbrock 
\& Dahari 1983) [Binette et al. (1990) suggested an even higher value for the 
intrinsic Balmer decrement in AGNs: \ha /\hb\ = 3.4]. The histogram 
has two main peaks showing a clear separation between strong \oiii$\lambda$5007 
objects (Seyfert 2s) and weak \oiii$\lambda$5007 objects (Liners). Although 
our sample is heterogeneous and incomplete, this suggests that there is 
no continuity between the two classes of objects. Heckman (1980) 
originally defined Liners as objects with \rliner\ $>$ 0.33; it seems that 
\rliner\ $>$ 0.25 would be a more realistic definition, as the observed 
distribution of this ratio really shows a minimum centered around this value. 

According to Ho et al. (1997a), the separation between the two principal 
ionization sources (young stars {\it vs.} AGNs) and between the two AGN 
excitation classes (Seyfert 2 {\it vs.} Liners) does not have sharp, 
rigorously defined boundaries. Fig. \ref{dd} shows that this is not the 
case. In fact, the three areas containing the \hii\ regions, the Seyfert 2s 
and the Liners are clearly separated; almost every ``transition object'' turns 
out to be ``composite'' when observed with sufficient resolution. 

Several authors had already suspected this to be the case. Kennicutt et 
al. (1989) and Ho et al. (1997c) have shown that the distribution of 
\hii\ nuclei in the \roiii\ {\it vs.} $\lambda$6583/ \ha\ plane parallels 
the disk \hii\ region sequence, the most striking feature being a clear 
offset between the two classes of objects, the \hii\ nuclei having larger 
\rnii\ ratios for the same excitation; this effect could be due to the 
presence of a weak active nucleus in many of these galaxies. 
Binette (1985) also suggested that mixed cases of starburst and 
Liner spectra might be relatively common, providing a possible interpretation 
for objects which have an unusually strong \roi\ ratio compared to \hii\ 
regions (NGC 3994, for example). Filippenko \& Terlevich (1992) 
suggested that Liners with weak \oi\ emission (\roi\ $<$ 1/6) might be powered 
by hot main-sequence stars; however, Ho et al. (1993a) showed that these 
objects are most probably ``composite''.

Ho et al. (1993b) reported the discovery of a non random trend in the 
dispersion of emission-line intensity ratios for Sey\-fert 2s. \roi\ and 
\rnii\ were found to be correlated with \roiii, suggesting the influence 
of a single underlying physical parameter -- the hardness of the ionizing 
continuum. Our data do not show these correlations, which could be artifacts 
due to the inclusion in the sample of ``composite'' spectra.

\begin{figure}
\resizebox{8.4cm}{!}{\includegraphics{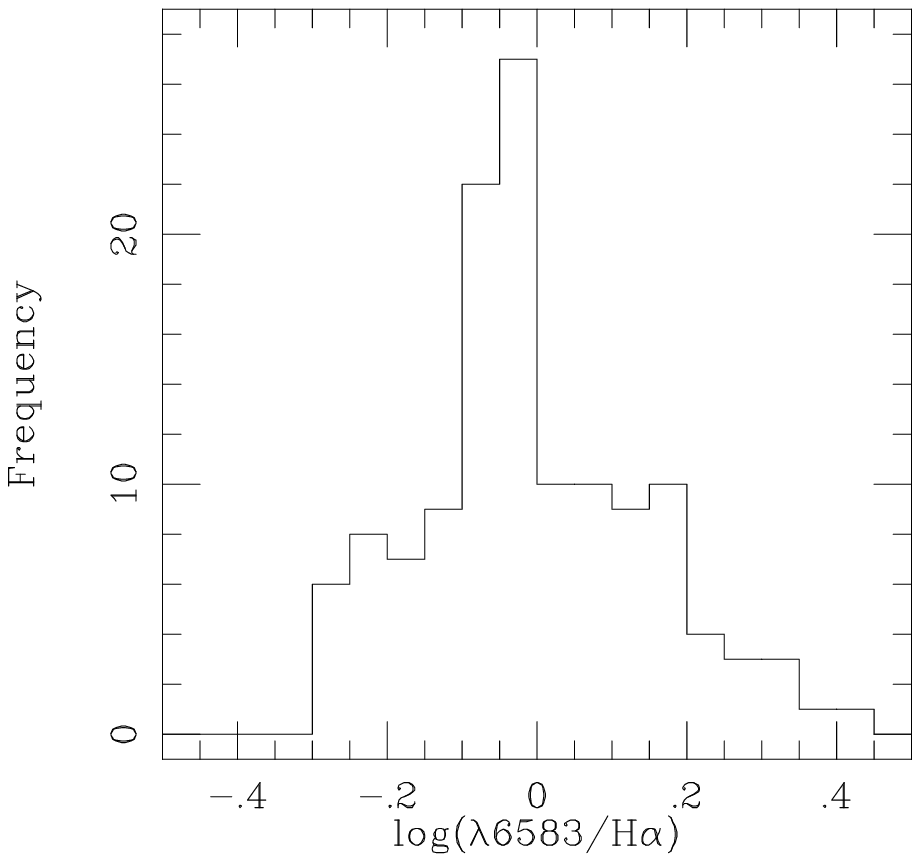}}
\caption{\label{histo_nii_ha}
Histogram of log($\lambda$6583/\ha) for the 131 Seyfert 2 galaxies 
plotted in Fig. \ref{dd}.}
\end{figure}

Examination of Fig. \ref{dd} shows that the points representative of Seyfert 2 
galaxies are not distributed at random in the region assigned to them. 
Figure \ref{histo_nii_ha} is the histogram of the 
quantity log(\rnii); it shows a 
sharp maximum at $\sim$ $-$0.05, with broad wings. Our sample of 
(131) Seyfert 2 galaxies is not complete in any sense and this could 
therefore be due to observational biases although this seems unlikely, as 
the \rnii\ ratio is not used for finding Seyfert 2 galaxies. We have no 
explanation for this fact.


\section{Discussion}

\subsection{The blue continuum in Seyfert 2 galaxies}

Koski (1978) and Kay (1994) found that all Seyfert 2 galaxies show 
an ultraviolet excess and 
weak absorption lines when compared with galaxies with no emission lines, 
indicating the presence of a blue featureless continuum. Boisson 
\& Durret (1986) and Vaceli et al. (1997) suggested that this continuum is 
a non-thermal power-law continuum. Kinney et 
al. (1991) argued that most of the Seyfert 2s in which a blue continuum has 
been observed are of type Sb or earlier, suggesting that it is truly 
associated with the Seyfert nucleus. Shuder (1981) 
showed that its strength and the \ha\ luminosity are 
strongly correlated suggesting that a direct physical connection exists between 
the two; studying a sample of 28 Seyfert 2s, Yee (1980) found that the 
\hb\ and continuum fluxes (rather than luminosities) are proportional over 
two orders of magnitude, with, however, a relatively large dispersion; but 
a number of those objects are now known to be Seyfert 1 galaxies. 

Martin et 
al. (1983) discovered that a small fraction of all Seyfert 2 galaxies have a 
highly polarized continuum. Subsequently, Antonucci \& Miller (1985), 
Miller \& Goodrich (1990) and Tran et al. (1992) showed that 
these objects harbour a hidden Seyfert 1 
nucleus, the observed polarized continuum arising from scattering of the 
nuclear continuum by dust or warm electrons. But most Seyfert 2s have very 
little polarization (Martin et al. 1983), much less than expected in the 
reflection model (Miller \& Goodrich 1990). 

On the other hand, Terlevich et 
al. (1990) showed that in Seyfert 2 galaxies, the IR \caii\ triplet is equal 
or, in some cases, higher than in normal elliptical galaxies, which is most 
naturally explained by the presence of young stars contributing heavily to 
the nuclear light at near-IR wavelengths. 

Heckman et al. (1995) used {\it International Ultraviolet Explorer (IUE)} 
spectra of 20 of the brightest type 2 Seyfert nuclei to build an ultraviolet 
template for this class; while the continuum was 
well detected in the template, there was no detectable broad line region (BLR), 
implying that no more than 20\%\ of the template continuum could be light 
from a hidden Seyfert 1 nucleus scattered by dust; they suggested 
that either most of the nuclei in their sample were ``pure'' type 2 Seyfert 
galaxies for which we have a direct view of the central engine and which 
simply lack of BLR, or that most of the observed ultraviolet 
continuum is produced by starbursts. From the absence of polarization of the 
continuum of most Seyfert 2 galaxies and of broad Balmer lines, 
Cid Fernandez \& Terlevich (1995) concluded that, most probably, this 
continuum was due to a population of young stars in the vicinity of the 
nucleus. Colina et al. (1997) obtained 
ultraviolet {\it HST} images of four nearby Seyfert 2 galaxies known to 
have circumstellar star-forming rings, providing direct empirical evidence 
that the UV flux emitted by these galaxies is dominated by radiation 
coming from clusters of young hot stars distributed along the star-forming 
ring. If similar rings are a common characteristic of Seyfert 2 
galaxies, the large {\it IUE} aperture would include both the Seyfert 2 nucleus 
and the rings for distances larger than 25 Mpc. Gonzalez Delgado 
et al. (1998) presented {\it HST} images and ultraviolet spectra of three 
Seyfert 2 nuclei (IC 3639, NGC 5135 and IC 5135); the data show the existence 
of nuclear starbursts (with absorption features formed in the photosphere of 
late O and early B stars) dominating the ultraviolet light. It is remarkable 
that, of the three observed galaxies, two (NGC 5135 and IC 5135) have a 
``composite'' nuclear emission spectrum, while the third (IC 3639), which has 
the largest UV nuclear flux (associated with the Seyfert nucleus) relative to 
the total UV flux, has a pure Seyfert 2 spectrum due to the relative weakness 
of the starburst emission component. 

We conclude that there is ample evidence for the presence of young, hot 
stars in the nuclear region of many Seyfert 2 galaxies. When the  
continuum is relatively bright, 
the associated \hii\ region could be strong enough to displace the 
object into the 
``transition'' zone in the diagnostic diagrams. 

AGNs are more frequent in 
early type galaxies while starbursts 
are more often found in late-type galaxies (V\'eron \& V\'eron-Cetty 1986; 
Ho et al. 1997b; Vaceli et al. 1997). It is therefore rather surprising to 
find almost systematically a population of young stars in Seyfert 2 galaxies; 
perhaps the nuclear activity triggers the star formation?

\subsection{Excitations and abundances in \ion{\it H}{\it II} nuclei and AGNs}

The \roiii\ and \rnii\ ratios are strongly correlated in \hii\ regions. 
Theoretical studies show that the hea\-vy-metal abundances 
change continuously along this 
sequence, a low \roiii\ ratio indicating a high metal abundance and a high 
\roiii\ ratio, a low metal abundance, with the heavy metal abundances changing 
from 1.5 Z$_{\sun}$ at the lower right of Fig. \ref{dd}a to 
0.25 Z$_{\sun}$ at the 
upper left (see for instance Dopita \& Evans 1986; Ho et al. 1997b). 
However, Stasi\'nska \& Leitherer (1996) have shown that most startbusts 
and \hii\ galaxies can be described as being produced by an evolving 
starburst with an universal initial mass function embedded in a gas cloud 
of the same metallicity. The emission line ratios depend mainly on two 
independent parameters: the age of the starburst and the metallicity. In 
this scenario, the \roiii\ ratio effectively changes with these two parameters 
and therefore is not a direct measurement of metallicity. 
The metallicity is strongly correlated with luminosity, luminous 
galaxies having higher metallicities; this correlation is also valid for 
elliptical galaxies, for which the metallicity is determined from absorption 
lines with [O/H] $\sim$ 1 at $M\rm_{B} = -$21 
(Salzer et al. 1989; Zaritsky et al. 1994).

AGNs are known to occur preferentially in high luminosity (Ho et al. 1997b), 
early-type (V\'eron \& V\'eron-Cetty 1986; Vacali et al. 1997) galaxies; 
they are therefore expected to have high metallicities. 
Indeed, the NLRs of active galactic nuclei have enhanced nitrogen abundances 
(Storchi-Bergmann \& Pastoriza 1989, 1990; Storchi-Bergmann et al. 1992; 
Schmitt et al. 1994). In these NLRs, [N/O] correlates with [O/H] in a 
manner identical to \hii\ regions in normal galaxies, with nuclear [O/H] 
and [N/O] values ranging from 1 Z$_{\sun}$ to 2 Z$_{\sun}$ (Storchi-Bergmann et 
al. 1996b). Storchi-Bergmann et al. (1996b,c) have determined the chemical 
composition of the \hii\ regions in the ring surrounding the nucleus of several 
AGNs, as well as in the nuclei; high metallicities were found ([O/H] 
$\sim$ 2 Z$_{\sun}$ and 
[N/O] $\sim$ 3 Z$_{\sun}$) both in the \hii\ regions and in the AGNs, 
these abundances being 
similar to those found in the nuclei of non-active galaxies with the same 
morphological type and absolute magnitude. Further work by 
Storchi-Bergmann et al. (1998) has shown that, in fact, oxygen abundances 
derived for Seyfert 2 nebulosities and neighbouring \hii\ regions (assuming 
that the emission lines in the active nucleus are due to photoionization 
by a typical active galactic nucleus continuum) are well correlated, while 
this is not the case for Liners. This suggests that the gas in AGNs 
and in the neighbouring \hii\ regions has the same origin and that the 
scatter observed in the Seyfert 2 region in the 
diagnostic diagrams, involving the \rnii\ ratio, is due to variations in the 
nitrogen abundance. In NGC 6300, in which \rnii\ = 3.4, the 
nitrogen abundance is estimated to be $\sim$ 5 Z$_{\sun}$.

We have seen that nuclear \hii\ regions and Seyfert 2 nebulosities, when 
appearing in the same galaxy, have the same high metallicity; as a 
result of their metallicity, the \hii\ regions have a low excitation, while 
the Seyfert 2 nebulosities have a high excitation. This explains why it is 
relatively easy to separate the two components in ``transition'' spectra.
 
\subsection{Objects with weak [\ion{\it N}{\it II}] lines}
\label{weak_NII_lines}

Figure \ref{dd} shows a small number of objects which 
have very weak \nii\ lines for Seyfert 2 galaxies; their \oi\ lines 
are however normal for this class of objects.

The first photoionization models invoked to explain the narrow emission lines 
in AGNs assumed a single density cloud. However, new observations quickly 
sug\-gested the presence of several emitting clouds, ruling out single 
component models. Most of the multicloud models first studied 
were such that the emitting 
gas, as a whole, was ionization-bounded and thus the \heii\,$\lambda$4686 
line intensity relative to \hb\ was 
determined by the hardness of the ionizing spectrum. In these models, the 
extreme values reached by the \rheii\ ratio are not 
well reproduced. A number of 
objects have \rheii\ of the order of 0.2 or more; such high values cannot be 
accounted for unless the line emitting clouds are matter-bounded 
(Stasi\'nska 1984). On the 
basis of a weak trend for the low excitation lines to become weaker as 
\rheii\ gets larger, Viegas-Aldrovandi (1988) and Viegas \& Prieto (1992) argued 
in favor of a model in which matter-bounded clouds are present; indeed, if the 
gas is not optically thick to all the ionizing continuum (i.e., is matter 
bounded), the H$^{+}$  emitting volume is smaller, but the He$^{++}$  volume 
is not, leading to a higher \rheii\ line ratio. Moreover, Viegas-Aldrovandi 
\& Gruenwald (1988) and Rodriguez-Ardila et al. (1998) showed that, for most 
AGNs, the observed low-excitation lines are better explained by matter-bounded 
mo\-dels with about 50\% of the \hb\ luminosity produced in 
ioniza\-tion-bounded clouds.

Storchi-Bergmann et al. (1996a) have obtained long-slit spe\-ctra of five 
active galaxies showing extended high excitation lines. At some positions, 
two of the objects (PKS 0349$-$27 and PKS 0634$-$20) show quite peculiar 
line ratios, with a strong \heii\,$\lambda$4686 line (\rheii\ $>$ 0.3) and 
weak \nii\ lines (that is, \rnii\ $<$ 0.3). In fact, there seems to be 
a correlation between \rnii\ and \rheii, weak \nii\ lines being associated 
with strong \heii\ emission, suggesting that very small \rnii\ ratios 
(as observed in the two above mentioned radiogalaxies) are not necessarily 
a signature of star-forma\-tion, but a natural consequence of having a 
region dominated by matter-bounded clouds (Binette et al. 1996, 1997). 
However, in the extranuclear regions of PKS 0349$-$278 in which strong 
\heii\,$\lambda$4686 and weak \nii$\lambda$6583 lines are observed, the 
\oi\ $\lambda$6300 line is also reduced (\roi\ $\sim$ 0.05), which is 
a natural consequence of the model (Viegas-Aldrovandi 1988), while in 
our sample of weak \nii$\lambda$6583 galaxies, we verify that the 
\oi\ $\lambda$6300 line is not weakened in most of the objects.

In Table \ref{nii_line_ratios} we give the list of known AGNs with relatively 
weak \nii\ lines (\rnii\ $<$ 0.45) with published values of the \rheii\ and 
\roi\ ratios. Three objects in this table (UM 85, MS 04124$-$0802 and Mark 
699) have both weak \nii\ lines (\rnii\ $<$ 0.20) and a strong \heii\ line 
($\lambda$4686/ \hb\ $>$ 0.30). 
In the last two, the \oi\ lines are also relatively weak (\roi\ $\le$ 0.05); 
these two objects could be dominated by matter-bounded clouds. Alternatively, 
in the other objects, the weakness of the \nii\ lines could be due to a 
selective under-abun\-dance of nitrogen. For a photoionized single cloud model 
with $U \sim$ 10$^{-2.5}$, Ferland \& Netzer (1983) predicted \rnii\ 
$\sim$ 1.0 for solar nitrogen abundances and $\sim$ 0.3 for nitrogen 
abundances $\sim$ 0.3 solar.

\begin{table}[h]
\begin{center}
\caption{\label{nii_line_ratios}
Known AGNs with weak \nii\ lines.}
\begin{flushleft}
\begin{tabular}{lllll}
\hline
Name  & \verb+  +Short & \underline{$\lambda$6583$\:$} 
& \underline{$\lambda$4686$\:$} & \underline{$\lambda$6300$\:$} \\
      & \verb+ +position & \verb+ + \ha  & \verb+ + \hb & \verb+ + \ha \\
\hline
UM 85            & 01\,04$+$06 & \verb+ +0.19 & \verb+ +0.37 & \verb+ +0.09  \\
MS 04124$-$0802	 & 04\,12$-$08 & \verb+ +0.10 & \verb+ +0.35 & \verb+ +0.05  \\
IRAS 04210$+$0400& 04\,21$+$04 & \verb+ +0.35 & \verb+ +0.21 & \verb+ +0.13  \\
3C 184.1	 & 07\,34$+$80 & \verb+ +0.22 & \verb+ +0.26 & \verb+ +0.07  \\
IRAS 11058$-$1131& 11\,05$-$11 & \verb+ +0.38 & \verb+ +0.17 & \verb+ +0.05  \\
SBS 1136$+$594	 & 11\,36$+$59 & \verb+ +0.10 & \verb+ +0.18 & \verb+ +0.11  \\
Mark 477 	 & 14\,39$+$53 & \verb+ +0.36 & \verb+ +0.13 & \verb+ +0.17  \\
Mark 699	 & 16\,22$+$41 & \verb+ +0.20 & \verb+ +0.34 & \verb+ +0.03  \\
\hline
\end{tabular}
\end{flushleft}
\end{center}
\end{table}

\subsection{Seyfert 2s and Liners}
 
It has been suggested by several authors (see for instance 
Ferland \& Netzer 1983; Shields 1992; Ho et 
al. 1993a) that in Seyfert 2s, as well as in Liners, the ionized gas is 
excited by a non-thermal continuum, the only differences being the value of 
the ionizing parameter which would be $\sim$ 10$^{-3.5}$ for Liners, and 
$\sim$ 10$^{-2.5}$ for Seyfert 2s. If this is the case, the discontinuity 
between Seyfert 2s and Liners is not easily understood. No reliable detection 
of the \heii\ line in {\it {bona fide}} Liners has been reported suggesting 
that there could be a serious problem with the picture of simply reducing 
$U$ in a standard power-law photoionization model predicting 
\rheii\ $>$ 0.15 (Viegas-Aldrovandi \& Gruenwald, 1990), as the weakness of 
\heii\ indicates that the continuum illuminating the NLR clouds must contain 
few photons more energetic than 
54.4 eV, the ionization potential of He$^{+}$ (P\'equignot 1984). Binette 
et al. (1996) proposed that the emission spectrum 
of Liners is due to ionization-bounded clouds illuminated by a ionization 
spectrum filtered by matter-bounded clouds hidden from view by obscuring 
material. In this case, the \heii\ emission is reduced (\rheii\ $<$ 0.01). 
However, a nearly total obscuration of the matter-bounded component must then 
be invoked in order to keep the emission from \heii\ at an acceptable low 
level, a scenario which seems to be rather unlikely to Barth et al. (1996).

\section{Conclusions}
We have shown that:
\begin{itemize}
\item{Nuclear \hii\ regions, Seyfert 2s and Liners lie in distinct, well 
separated regions in the log(\roiii) {\it {vs.}} log($\lambda$6583/ \ha) 
and log(\roiii) {\it {vs.}} log(\roi) diagrams. The\-re is no continuity 
between Liners and Seyfert 2s, with an apparent deficit of objects with 
\rliner\ = 0.25.}
\item{A number of objects have ``transition'' spectra, falling outside the 
regions assigned to the three types of emission nebulosities. They probably 
all have a ``composite'' spectrum.}
\item{We have isolated a class of Seyfert 2 galaxies with weak \nii\ lines. 
This weakness could be due to an under-abun\-dance of nitrogen or to the 
presence of matter-bounded clo\-uds in these objects.}
\end{itemize}
 

\begin{acknowledgements} 
This research has made use of the NASA / IPAC 
extragalactic database (NED) which is ope\-rated by the Jet Propulsion 
Laboratory, Caltech, under contract with the National Aeronautics and Space 
Administration. A.\,C. Gon\c{c}al\-ves acknowledges support from the
{\it Funda\c{c}\~ao para a Ci\^encia e a Tecnologia}, Portugal, during the 
cou\-rse of this work (PhD. grant ref. PRAXIS XXI/BD/5117/95).
\end{acknowledgements}


\end{document}